\documentclass[12pt]{article}

\usepackage{amssymb}
\usepackage[mathscr]{euscript}
\usepackage{graphicx}
\usepackage{yfonts} 
\usepackage[all,knot]{xy}
\usepackage{multicol}
\usepackage[percent]{overpic}

\usepackage[T1]{fontenc}
\usepackage{amsmath}

\PassOptionsToPackage{}{amsmath}
\RequirePackage{amsthm}
\RequirePackage{amsmath}
\RequirePackage{latexsym}
\RequirePackage{amssymb}
\RequirePackage{amscd}
\RequirePackage{epsfig}
\RequirePackage{graphics}
\RequirePackage{ifthen}
\RequirePackage{varioref}

\usepackage{tikz} 
\include{tikzcolor}
\usetikzlibrary{matrix,decorations.pathreplacing}

 \usepackage{etex}

\newtheorem{prop}{Proposition}[section]

\newtheorem{defi}{Definition}[section]
\newtheorem{lemm}{Lemma}[section]

\newcommand{\iN}{\hbox{ {\leaders\hrule\hskip.2cm}{\vrule height .22cm} }}

\newcommand{\MM}{\textsf{m}}

\newcommand{\R}[1][]{\ensuremath{{\mathbb{R}^{#1}} }}

\newcommand{\m}{\textsc{m}}
\newcommand{\z}{\textsf{z}}

\newcommand{\pp}{\textsf{p}}

\newcommand{\x}{\textsf{x}}
 \newcommand{\y}{\textsf{y}}

\newcommand{\ypoint}{\dot{\textsf{y}}}

\newcommand{\mpoint}{\dot{\textsf{m}}}

\newcommand\Proof{{{\textsf{\textcolor{warmblack}{Proof}}}}}

\newcommand{\ST}{\cal X}
\newcommand{\po}{\overset{\hbox{\tiny{$\mathfrak{0}$}}}{{{\psi}}}}
\newcommand{\poo}{\overset{\hbox{\tiny{$\mathfrak{1}$}}}{{{\psi}}}}

\newcommand{\Xio}{\overset{\hbox{\tiny{$\mathfrak{0}$}}}{{{\Xi}}}}
\newcommand{\Xioo}{\overset{\hbox{\tiny{$\mathfrak{1}$}}}{{{\Xi}}}}

\newcommand{\etao}{\overset{\hbox{\tiny{$\mathfrak{0}$}}}{{{\eta}}}}
\newcommand{\etaoo}{\overset{\hbox{\tiny{$\mathfrak{1}$}}}{{{\eta}}}}
\newcommand{\nbetao}{\overset{\hbox{\tiny{$\mathfrak{0}$}}}{{{\eta}}}}
\newcommand{\nbetaoo}{\overset{\hbox{\tiny{$\mathfrak{1}$}}}{{{\eta}}}}
\newcommand{\bfpo}{\overset{\hbox{\tiny{$\mathfrak{0}$}}}{{{\psi}}}}
\newcommand{\bfpoo}{\overset{\hbox{\tiny{$\mathfrak{1}$}}}{{{\psi}}}}
 
\newcommand{\momo}{\overset{\hbox{\tiny{$\mathfrak{0}$}}}{{{p}}}}
\newcommand{\momun}{\overset{\hbox{\tiny{$\mathfrak{1}$}}}{{{p}}}}

\newcommand{\etazero}{\overset{\hbox{\tiny{$\mathfrak{0}$}}}{{{\eta}}}}
\newcommand{\etaun}{\overset{\hbox{\tiny{$\mathfrak{1}$}}}{{{\eta}}}}   
\newcommand{\PB}{\cal P}
\newcommand{\g}{\mathfrak{g}}
\newcommand{\p}{\mathfrak{p}}

\newcommand{\G}{\mathfrak{G}}

\usepackage{color}
\definecolor{warmblack}{rgb}{0.0, 0.26, 0.26}
\definecolor{slategray}{rgb}{0.44, 0.5, 0.56}
\definecolor{darkjunglegreen}{rgb}{0.1, 0.14, 0.13}
\usepackage[colorlinks,linkcolor=slategray,citecolor=darkjunglegreen,urlcolor=darkjunglegreen]{hyperref}
  
\title{10--plectic formulation of        gravity   \\ and   Cartan connections}
\author{Dimitri VEY\footnote{
\textsf{dim.vey@gmail.com}} }

\begin{document}
\maketitle

\emph{Abstract} ---  We  give a Hamiltonian formulation of  
       Weyl--Einstein--Cartan   gravity which is   covariant from the viewpoint of the geometry of the principal fiber bundle.  
     The connection is   represented by a  $1$-form with values in the Poincar\'{e} Lie algebra, which is defined   on the total space of the orthonormal
frame bundle fibered over the space-time.  
Within the $10$-plectic framework we discover  that  
 the local equivariance property of the  Cartan
connection is  a consequence of  the Hamilton
equations.

\section{Introduction}

 One of the guiding ideas rooted in this paper is that the {\it{physical laws}} (in particular General Relativity)  are independent of the point of the space-time manifold {\bf{and}}   the pseudo-orthonormal frame  (or reference frame)   in which they are expressed. This viewpoint is  adopted in   the  approach to field theory based on the {{space of  reference frames}}, which is developed  by   Toller \cite{Toller1977,Toller1978,Toller1996}. This standpoint echoed  the  work initiated  by   Lur{\c c}at \cite{lurcat}     towards  a  {{Quantum Field Theory on the Poincar\'{e} group}}  (see also   \cite{Lurcat2002a,Varlamov}).  The original motivation of Lur\c{c}at was to put  the dynamical role of spin in the foreground.        Accordingly,   in  {{gauge theories of gravitation}}   the local model of Minkowski space is replaced  by  the Poincar\'{e} group \cite{Hehl2013,Sardanashvily16}.     We  present a Hamiltonian formulation of first order    gravity which is   covariant from the viewpoint of the geometry of the principal fiber bundle,  {i.e.}   
 which does not depend on choices  of space-time coordinates {\em{nor}} on the trivialization of the principal bundle. {For a formulation which is covariant from the viewpoint of  the geometry of space--time only, we refer to    \cite{Vey2015} and references therein.}   We work with    the multisymplectic approach,  which  goes back to the discovery of generalizations of the Hamilton equations for variational problems with several variables  by  Volterra  
\cite{volterra1,volterra2}. These ideas were   first    developped in the thirties     \cite{caratheodory,Donder01,Lepage,Weyl} 
and later in the seventies of the last century  \cite{Garcia00,Garcia74,HGSS,Kijowski1974,KS1,JKWMT,Roman}. 
The multisymplectic   theory  has many recent developments  (see e.g.  \cite{Cantrijn1,forgergomes,Gotay91a,heleinleeds,hk2}) 
 The 
basic concept is the notion of a multisymplectic   $(m+1)$-form ${{\pmb{\omega}}} $ on a smooth manifold $\mathcal{N}$, where $m$ corresponds  to the number of independent variables. The form ${{\pmb{\omega}}} $ is  always  closed and one often assumes that it is non degenerate: $\forall \xi \in {{\Gamma}}({\cal N},T{\cal N})$,  $\xi\iN  {{\pmb{\omega}}} = 0 \Rightarrow \xi = 0$.
 The solutions of the      Hamilton--Volterra--De Donder--Weyl ({HVDW})     equations are given  by oriented $m$-dimensional submanifolds ${\pmb{\gamma}}$ of $\mathcal{N}$ which satisfy the condition that, at any point $\MM\in \mathcal{N}$, there exists a basis $(X_1,\cdots,X_m)$ of $T_\MM{\pmb{\gamma}}$ such that $X_1\wedge \cdots \wedge X_m\iN  {{\pmb{\omega}}}  = (-1)^md H$, where   $H:\mathcal{N}\longrightarrow \Bbb{R}$  is a Hamiltonian function.  Analogously one can replace ${{\pmb{\omega}}} $ by its restriction to the level set $H^{-1}(0)$
and describe the solutions as the submanifolds ${\pmb{\gamma}}$ of $H^{-1}(0)$ such that
$X_1\wedge \cdots \wedge X_m\iN {{\pmb{\omega}}}   = 0$ everywhere   and $\beta^{(m)}(X_1\wedge \cdots \wedge X_m  )\neq 0$, where $\beta^{(m)}$ is some volume $m$-form, 
see  \cite{heleinleeds}.

\emph{Organisation of the paper} ---  
In  Section \ref{space-time-dynamics} we  describe  the space-time dynamical fields $(e,A)$ for  Weyl--Einstein--Cartan (WEC) gravity. The multisymplectic formulation of the WEC  functional   ${{{ \mathcal{S}}}}_{\hbox{\tiny{\sffamily WEC}}}  [e,A] $  is delicate    because    the  dynamical fields $(e,A)$ are the components   of a Cartan connection (see \cite{Cartan,Cartan2,Ehresmann,Marle}) and  subject to gauge invariance.  
In Section  \ref{subsubsec:lift-principal-bundle} we lift the connection defined
on some manifold ${\ST}$ representing the space-time to the principal bundle $\mathcal{P}$ over $\ST$ with
structure group the Lorentz group.  The connection is   represented by a 1-form $(\alpha,\omega)$ on $\mathcal{P}$ with value in the Poincar\'{e} algebra  $\p = \g\oplus \textswab{t}$,  which satisfies  {\it normalization}  
 and    {\it equivariance} 
 hypotheses, see Equations \eqref{normalisationStandard}.  
 Although {\it a priori} mandatory  the equivariance condition has the shortcoming of being a {non holonomic} constraint,  {i.e.} on  the first order derivatives of the field. 
 
   The geometrical background for the  $10$-plectic formulation of WEC gravity is presented in Section \ref{sec:MFP}.  The covariant configuration space is    ${\cal Z} := \mathfrak{p}\otimes T^*\mathcal{P}$, the space of $\p$-valued $1$-forms over $\mathcal{P}$.    Section \ref{geo-back} is dedicated to present  the WEC $10$-form,  the first   order jet bundle  $ J^1{\cal Z}$   and   the De Donder--Weyl (DW) bundle $ \Lambda^{10}_{\mathfrak{1}}T^{\ast}\mathcal{\mathcal{Z}}$.
  In the following,   we compute the Legendre transform for the WEC  action    by treating connections as normalized and equivariant $\p$-valued 1-forms on $\mathcal{P}$ (see Section \ref{LC-1}).  
 We find that the natural multisymplectic manifold can be built from the vector
 bundles $\p\otimes T^\ast\mathcal{P}$ and $\p^\ast\otimes \Lambda^{8}T^*\mathcal{P}$ over $\mathcal{P}$, where
 $10$ is the dimension of $\mathcal{P}$,
 $\p$ is the  structure Lie algebra and $\p^\ast$ its dual vector space. These vector
 bundles are endowed with a canonical $\p$-valued 1-form $\eta $ and a canonical  
 $\p^\ast$-valued $8$-form $\psi$ respectively (see also \cite{HeleinVey01a}).   Then the multisymplectic
 manifold corresponds  to the total space of the vector bundle
$
 {\mathcal{M}} := \R\oplus_\mathcal{P} (\p\otimes^\textsc{n} T^*\mathcal{P})\oplus_\mathcal{P}(\p^*\otimes \Lambda^{8}T^*\mathcal{P})$, 
  equipped with the $10$-form $  \theta^{(10)} = \varsigma \beta^{(4)} \wedge \gamma^{(6)}   +  \psi \wedge  (d \eta + \eta \wedge \eta),$
 where $\varsigma$ is  a coordinate on $\R$,   $\beta^{(4)}\wedge \gamma^{(6)}$ is the volume form on $\mathcal{P}$ and  
   $\p\otimes^\textsc{n} T^\ast\mathcal{P} \subset \p\otimes T^*\mathcal{P} $ is the subbundle  of   normalized forms. 
 
 Finally,  the DW formulation of the Hamilton 
  equations is given  in Section \ref{sec:HVDW-equations}. 
  Any  solution  of the Hamilton  equations is      given by      a  
 $10$-dimensional submanifold of ${\cal N}$, more precisely a section $\phi$ of ${\cal N}$ over ${\PB}$. In Section \ref{ole}, we compute the  $11$-plectic form  ${\pmb{\omega}}:=d \theta^{(10)}$.  In   Section \ref{WEC--HVDW-equations},  we finally discover   that         the       dynamical  equations    constrain  the $\p$-valued 1-forms to be equivariant, see Proposition \ref{HVDW-001}. In addition,  the formalism   yields     Einstein--Cartan  type     equations: 
  \begin{equation}\label{ECsystem} 
 \left\{
 \begin{array}{lcl}
   {G}{^b}_a & = & \frac{1}{2} \rho_j\cdot p{_a}{^{bj}}\\
    {T}{^a}_{cd} & = & - \left( \textsf{h}_{de}\delta^a_{a'}\delta^{c'}_c
 + \frac{1}{2}\delta^{c'}_{a'}(\delta^a_d\textsf{h}_{ce}
 - \delta^a_c\textsf{h}_{de})\right) \rho_j\cdot p{_{c'}}{^{ea'j}} ,
 \end{array}
 \right.
 \nonumber
\end{equation}
where   $  {G}{^b}_a$ is  the Einstein  tensor and   $  {T}{^a}_{cd} $  is   the  torsion  tensor,  see  Section \ref{Einstein-Spin} in the Annex. In addition, $\rho_j$ is a left invariant vector field acting on the multimomenta coordinates   $p{_a}{^{bj}}$ and  $p{_{a}}{^{bcj}}$ 
 which are   given in  Proposition \ref{HVDW-002}.  
 
 The approach  of  curved space-time by  {{crystallization of liquid fiber bundles}}, 
which is developed elsewhere (see \cite{HeleinVey01a}),  is  more fundamental and includes    the   one   given in this paper as a peculiar case. Hence,  {Proposition \ref{HVDW-002}
  reproduces  partially the results obtained  in the broader context of {\it{liquid fiber   bundles}}, where the Hamilton equations contain in addition     non homogeneous Maxwell type equations
(see Equations (\textcolor{slategray}{88}) in \cite{HeleinVey01a})}.  In the former,  no  {\it a priori}  hypotheses are  given to  settle  the structure   of  the principal  fiber bundle. We refer also to various works of Hélein \cite{helein20,helein22,helein25} for a deep presentation  of  gauge and gravity theories on a dynamical principal bundle, including Kaluza-Klein theories, Einstein--Cartan, Palitini theories. We also refer to \cite{Pierard}, which adresses the question of the inclusion of the Dirac fields for a complete gravity theory.

\noindent
\emph{Aknowledgements:} ---       I am indebted to  F.       H\'{e}lein   for  being    a    co--architect     and co--explorer    of this work,  both       at its origins and    developments; I thank him for comments and  corrections      on   preliminary  variations of the paper.

  \section{Weyl--Einstein--Cartan}\label{sec:WEC}

\noindent 
Let    $\vec{\mathbb{M}}$ be the  Minkowski vector space endowed with the Minkowski metric $\textsf{h}$. We  fix a pseudo-orthonormal basis $({E}_a)_{0\leq a \leq 3}$  of $(\vec{\mathbb{M}},\textsf{h})$. In addition,  $\mathfrak{T}$ is the Abelian Lie group of translations on   $\vec{\Bbb{M}}$,   $\mathfrak{t}$ its  trivial Lie algebra, with basis $({\mathfrak{t}}_a)_{0 \leq a \leq 3}$ and $\mathfrak{t}^*$ the dual of $\mathfrak{t}$ with basis $({\mathfrak{t}}^a)_{0 \leq a \leq 3}$.   We denote by     $\G$     the Lorentz group $SO(3,1)$,  $\g$ its    Lie     algebra,  {i.e.}  $\g :=  {\bf so}(3,1)$ and $\g^{*}$ the dual of $\g$. We denote by  $({\textswab{l}}_{j})_{1\leq j\leq6} $ a basis of $\g$  and $({\textswab{l}}^{j})_{1\leq j\leq6} $ a  basis     of and $\g^*$, respectively. Finally,  $\mathfrak{G}\ltimes \mathfrak{T}$ is the        Poincar{\'e}  group  $ISO(3,1):= SO(3,1) \ltimes \mathfrak{T} $,  $\p := \g \oplus \mathfrak{t}$   its Lie algebra and $\p^{*}$  is the  dual of $\p$. We fix some basis  $(\mathfrak{l}_A)_{0\leq A\leq 9} = ({\mathfrak{t}}_0,\cdots,{\mathfrak{t}}_3, \mathfrak{l}_1,\cdots,\mathfrak{l}_6)$ and $(\mathfrak{l}^A)_{0\leq A\leq 9} = ({\mathfrak{t}}^0,\cdots,{\mathfrak{t}}^3, \mathfrak{l}^1,\cdots,\mathfrak{l}^6)$   of ${\p}$ and   $\p^*$, respectively.

 \subsection{Space-time dynamics}\label{space-time-dynamics} 
 
   In WEC formulation of gravity, dynamical fields    can be defined locally as being pairs $(e,A)$, where $e = (e^0, e^{1}, e^2, e^3)$ is a moving coframe on ${\cal X}$  and $A$ is a $\g$-valued connection $1$-form on ${\cal X}$. We set a volume $4$-form  $e^{(4)} := e^0\wedge e^1\wedge e^2\wedge e^3$ and  $e_{ab}^{(2)}:=  ( \frac{\partial}{\partial e^{a}}  \wedge \frac{\partial}{\partial e^{b}}   ) \iN e^{(4)}$. The WEC  action     reads 
   \begin{equation}
{{{ \mathcal{S}}}}_{\hbox{\tiny{\sffamily WEC}}}  [e,A] = \int_\mathcal{X}e_{ab}^{(2)} \wedge F^{ab}
 = \int_\mathcal{X} u^{ab}_{i}  e_{ab}^{(2)} \wedge F^i,
 \end{equation}
where  $F := d A+A\wedge A$ is the curvature form and $F^{cd}:= \textsf{h}^{dd'}F{^c}_{d'}$. {Elsewhere,  this action is  termed    the {{\guillemotleft~Palatini~\guillemotright}} action functional, which   is    inexact     \cite{Ferraris81}.}   
 \\
 \\
\noindent {\textsf{Torsion and curvature}} ---
The  torsion  and  curvature $2$-forms, which are denoted by   $T^{a} $ and  $F^{ab}$, are      related to the dynamical field $(e,A)$ by the   Cartan structure equations:
$ T^{a}   = d e^{a} + A^{a}{}_{b} \wedge e^{b} $ and 
$    F^{a}{}_{b} = d A^{a}{}_{b} + A^{a}{}_{c} \wedge A^{c}{}_{b} $, respectively. We    introduce the 
  torsion tensor $T{^a}_{cd}$   such that  $ T^{a}  = \frac{1}{2}T{^a}_{cd}e^{cd}  =  \frac{1}{2}    T^{a}_{\mu\nu}   \beta^{\mu\nu} $, where  the components $T^{a}_{\mu\nu} = T^{a}{}_{cd}  e^{c}_{\mu} e^{d}_{\nu}  $ are given by  
 $
 {T}_{\mu\nu}^{a}   =   
 \partial_\mu e_{\nu}^{a} - \partial_\nu e_{\mu}^{a}  + A^{a}_{\mu c} e_{\nu}^{c}  -  e^{c}_{\mu} A_{\nu c}^{a}
 $.
In addition, the curvature tensor  $F{^a}_{bcd}$ is  such that $ F{^a}_b =  \frac{1}{2} F{^a}_{bcd}e^{cd}  =   {F}^{ab}{}_{cd} e^{c}_{\mu} e^{d}_{\nu} $, where    the components  are written as   $
 {F}_{\mu\nu}^{ab}  
 = \partial_\mu A_{\nu}^{ab} - \partial_\nu A_{\mu}^{ab}  + A^{a}_{\mu c} A_{\nu}^{cb}  -  A^{a} _{\nu c} A_{\mu}^{cb}
 $. 
\\
\\
\noindent {\textsf{Ricci and Einstein tensors}} ---  We denote by   $ \hbox{Ric}_{ab} :=  F^{a'}{}_{aa'b}  = {\textsf{h}}_{ab'} F^{a'b'}{}_{a'b} $ the Ricci  tensor.  Then, 
%
  $G_{ab} = \hbox{Ric}_{ab} -  \frac{1}{2} \textsf{h}_{ab} \hbox{S}$ is the  Einstein tensor, 
where $\hbox{S}$ is the scalar curvature, which is given by    
$\hbox{S}=  {\textsf{h}}^{ab} \hbox{Ric}_{ab} = {\textsf{h}}^{ab}  {\textsf{h}}_{ab'} F^{a'b'}{}_{a'b}  =   F^{a'b}{}_{a'b} $.

  \subsection{Bundle dynamics}\label{subsubsec:lift-principal-bundle}

In addition, the   WEC  functional      is invariant by gauge transformations
of the form
$ (e,A)\longmapsto (g^{-1}e,g^{-1}dg + g^{-1}Ag)$, 
which are written  in indices as 
$
e^a\longmapsto (g^{-1}){^a}_{a'}e^{a'}$ and $
  A{^a}_b\longmapsto (g^{-1}){^a}_{a'}dg{^{a'}}_b + (g^{-1}){^a}_{a'}A{^{a'}}_{b'}g{^{b'}}_b
$, 
where $g:\mathcal{X}\longrightarrow \mathfrak{G}$. In order to fully consider the gauge invariance, 
  we   now   lift the theory to the  total space of 
%
%
the  principal fiber  bundle   $({\PB} ,   {\ST} , \pi , \G )$, where  ${\ST}$ is the base space,    
 $ {\PB} $ is  the total space, $\G$ is the Lorentz structure group and  $\pi_{{\ST}}:{\PB} \rightarrow  {\ST}$ is the fibration map. We assume that $\G$ is acting on the right on ${\PB}$:
\[     \left. 
 \begin{array}{cccc} 
{R}{}_g : & {\PB}  \times \G & \longrightarrow &{\PB}  \\
& (\z,g) & \longmapsto & \z\cdot g = {R}{}_g({\z})
  \end{array}\right.
 \]
This induces an infinitesimal action of $\g$, to any $\xi\in \g$,
we associate the vector field $\rho_\xi(\z) = \z \cdot \xi$  on ${\PB}$ defined by:
$
\forall \ {\z} \in {\PB}, \forall \xi\in \g, 
\rho_\xi(\z):=  {d}/{d t}(\z\cdot e^{t\xi})|_{t=0}$.
 For any $\z \in {\PB}$ the orbit of the $\G$ action containing $\z$ is the fiber ${\PB}_{\x}$,
where $\x = \pi_{{\PB}} (\z)$. The tangent vector subspace to ${\PB}_{\x}$ at ${\z}$ is   the  vertical subspace   ${{V}}_{\z} {\cal P} := {\hbox{ker}} d(\pi_{{\ST}})_{\z}$ and  is isomorphic to the Lie
algebra $\g$ of $\G$.    By choosing  a section ${\sigma}: {\ST}\rightarrow \mathcal{P}$ which induces
a trivialization $\z = {\sigma}(\x)\cdot g \simeq (\x,g)$,    we set    
$
 \partial_{\mu}(\z):=  d ({R}{}_g\circ {\sigma})_{\x}(\partial_{\mu}({\x}))\simeq 
\partial_{\mu}(\x)\cdot g,
   \hbox{for }\mu = 0,\cdots, 3
$, 
where  $({{\partial}}_\mu)_{0 \leq \mu\leq 3}$  is a    moving frame  on $ {\ST}$.  We  consider also  the   family of independent  tangent vector fields  $(\rho_i)_{1\leq i \leq6}$ on $\mathcal{P}$ induced by the right action of $u_i$ on $\mathcal{P}$, which, at every point
$ {\z}\in \mathcal{P}$, spans the vertical subspace ${V}_{\z} {\PB}$.
Then $(\partial_{\mu},\rho_i)_{0\leq \mu\leq 3 , 4 \leq i \leq 9}$ is a moving frame on $\mathcal{P}$.
The dual  frame  $(dx^{\mu}  , \gamma^{i} )_{0\leq \mu\leq 3 , 4 \leq i \leq 9}$  is  the family
of sections of $T^\ast\mathcal{P}$ such that $ dx^{\mu}(\partial_\nu) = \delta^\mu_\nu$ and $\gamma^i(\rho_j) = \delta^i_j$. 

To picture geometrically the gauge invariance we   lift
the variational problem on the  total space $\mathcal{P}$ of the principal bundle of orthonormal frames.   We  represent each pair $(e,A)$
by a pair of 1-forms $(\alpha,\omega)$
on $\mathcal{P}$ with values in $\mathfrak{p}$,  {i.e.}   
$\alpha$ takes values in $\mathfrak{t}$ and $\omega$ takes values in $\mathfrak{g}$.    However,  $(\alpha,\omega)$ needs to  satisfy the following  {normalization} and {equivariance} hypotheses:
 \begin{equation}\label{normalisationStandard}
\left\{
\begin{array}{rcl}
\displaystyle    \rho_i\iN\alpha & = &   \displaystyle   0 
\\
\displaystyle   \rho_i\iN  \omega  & = &   \displaystyle    u_i,
   \end{array}
\right.
 \quad \quad   \quad\quad
\left\{
\begin{array}{rcl}
\displaystyle     L_{\rho_i}\alpha + u_i\cdot \alpha   & = &   \displaystyle    0\\
\displaystyle     L_{\rho_i}\omega + [u_i,\omega]   & = &   \displaystyle    0,
   \end{array}
\right.
\end{equation}
where $L_{\rho_i}$ is the Lie derivative with respect to a vector field $\rho_i$.
\noindent
We can lift the
action $ \mathcal{S}_{\hbox{\tiny{\sffamily WEC}}} [e,A]$
to a functional on the space of $\mathfrak{p}$-valued 1-forms $(\alpha,\omega)$ by setting:
\begin{equation}\label{WEC-action-total}
 \widehat{\mathcal{S}}_{{\hbox{\sffamily{\tiny{WEC}}}}}[\alpha,\omega] = \int_\mathcal{P}
\alpha^{(2)}_{ab}\wedge \Omega^{ab}\wedge \gamma^{(6)}
= \int_\mathcal{P}u^{ab}_i\alpha^{(2)}_{ab}\wedge \Omega^i\wedge \gamma^{(6)}.
\end{equation}
where
 $\alpha^{ab}= \alpha^{a} \wedge \alpha^b$. By setting  $\alpha{}_{ab}^{(2)}:=  \left(  \frac{\partial}{\partial\alpha{}^a}\wedge   \frac{\partial}{\partial \alpha{}^b} \right) \iN\alpha{}^{(4)}   
 $, 
$\Omega:= d\omega+\omega\wedge \omega$,
$\Omega^{ab}:= \Omega{^a}_{b'}\textsf{h}^{bb'}$
and $\gamma^{(6)}:= \gamma^1\wedge \cdots \wedge \gamma^6$. 
Then critical points of ${\mathcal{S}}_{{\hbox{\sffamily{\tiny{WEC}}}}} [e,A]$ correspond to critical points of $\widehat{\mathcal{S}}_{{\hbox{\sffamily{\tiny{WEC}}}}}[\alpha,\omega]$
under the constraints (\ref{normalisationStandard}). 

For any $\mathfrak{p}$-valued 1-form
$(\alpha,\omega)$ on $\mathcal{P}$
which satisfies (\ref{normalisationStandard}) 
 and for any local section
$\sigma:\mathcal{X}\longrightarrow \mathcal{P}$, we obtain a pair $(e,A)$ on $\mathcal{X}$ simply by setting
$e=\sigma^*\alpha$ and $A=\sigma^*\omega$. Conversely, given a pair $(e,A)$ on $\mathcal{X}$ and a local section
$\sigma:\mathcal{X}\longrightarrow \mathcal{P}$, this provides us with a local trivialization
$
  {\tau}:  \mathcal{P}   \longrightarrow   \mathcal{X}\times \mathfrak{G} :  \textsf{z}   \longmapsto   (\textsf{x},g)
$, 
where $(\textsf{x},g)$ is such that $\textsf{z} = \sigma(\textsf{x})\cdot g$. We  
associate to $(e,A)$ a $\mathfrak{p}$-valued 1-form $(\alpha,\omega)$ on $\mathcal{P}$
which satisfies (\ref{normalisationStandard}),  
    given by $\alpha = {\tau}^*(g^{-1}e)$
and $\omega = {\tau}^*(g^{-1}Ag+g^{-1}dg)$.
    \begin{equation}\label{in-a-trivialization}
(    \alpha , \omega )   = (     {{g}} {}^{-1} {e} ,  {{g}} {}^{-1} d {g} + {g} {}^{-1}   {A}   {g}) \quad  \iff \quad (\alpha,\omega)  = (0,g^{-1}dg) + \hbox{Ad}_{g^{-1}} (e,A),
\end{equation}
where $(e,A)$ is a $\mathfrak{p}$-valued 1-form whose coefficients depend only on the
$x$ variables.     

\noindent   In particular,     by using the representation  $ {{\omega}} =   g^{-1} d g + g^{-1} A g 
$  and    $\alpha = g^{-1} e$, then we obtain   $    d {\omega} + {\omega}\wedge {\omega} = g^{-1}( d {A} +{A} \wedge {A})g $ 
 and $    d  {{\alpha}} +   {{\omega}} \wedge {{\alpha}} = g^{-1} \left( d e + {A} \wedge e \right)  $.

         \section{The $10$-plectic   formulation}\label{sec:MFP}
         
  \subsection{Geometrical background}\label{geo-back}
The covariant  configuration space is the $110$-dimensional  vector  bundle    $\mathfrak{p}\otimes T^*\mathcal{P}$
over $\mathcal{P}$, whose fiber at point $\textsf{z}\in \mathcal{P}$ is the tensor product
$\mathfrak{p}\otimes T^*_\textsf{z}\mathcal{P}$.  {Note that ${\hbox{dim}} ({\PB}) = 10$  and   $ {\hbox{dim}} (\p \otimes T^{\ast}_{\z}{\PB}) = {\hbox{dim}}(\p) \cdot {\hbox{dim}}(T^{\ast}_{\z}{\PB}) = 100.$} A point in $\mathfrak{p}\otimes T^*\mathcal{P}$ will be denoted by $(\textsf{z},\textsf{y})$,
where $\textsf{z}\in \mathcal{P}$ and $\textsf{y}\in\mathfrak{p}\otimes T^*_\textsf{z}\mathcal{P}$.

\subsubsection{Weyl--Einstein--Cartan $10$-form}

We consider the canonical $\mathfrak{p}$-valued 1-form $\eta$
(a section of $\mathfrak{p}\otimes T^*(\mathfrak{p}\otimes T^*\mathcal{P})$) defined by:
\[
 \forall (\textsf{z},\textsf{y})\in \mathfrak{p}\otimes T^*\mathcal{P},
\forall v\in T_{(\textsf{z},\textsf{y})}(\mathfrak{p}\otimes T^*\mathcal{P}),\quad
\eta_{(\textsf{z},\textsf{y})}(v) = \textsf{y}(d\pi_{(\textsf{z},\textsf{y})}(v)),
\]
where $\pi  :
\mathfrak{p}\otimes T^*\mathcal{P}\longrightarrow \mathcal{P}$ is the canonical projection
map. This $\mathfrak{p}$-valued 1-form can be decomposed as
$\eta = \mathfrak{l}_A\eta^A$, where each $\eta^A$ is a 1-form on $\mathcal{P}$.  Any pair $(\alpha,\omega)$ as considered in   previously   is a section of
$\mathfrak{p}\otimes T^*\mathcal{P}$ over $\mathcal{P}$. In the following we identify such a pair with
a map $\varphi$ from $\mathcal{P}$ to the total space of $\mathfrak{p}\otimes T^*\mathcal{P}$  such that
$\pi\circ \varphi(\textsf{z}) = \textsf{z}$, $\forall \textsf{z}\in \mathcal{P}$, by letting
 \begin{equation}\label{identification-forme-section}
  (\alpha,\omega) = \varphi^*\eta .
 \end{equation}

\noindent We introduce the following coordinates on $\mathfrak{p}\otimes T^*\mathcal{P}$.
   $(z^I)_{1\leq I\leq 10}$ are local coordinates on $\mathcal{P}$; thus they provide us with
 locally defined functions $z^I\simeq z^I\circ \pi$ on $\mathfrak{p}\otimes T^*\mathcal{P}$.   In a given   trivialization  $
  {\tau}:  \mathcal{P}   \longrightarrow   \mathcal{X}\times \mathfrak{G} :  \textsf{z}   \longmapsto   (\textsf{x},g)
$, we denote  by $(x^{\mu},g{})$ the coordinates functions for a point ${\z}\in  {\PB}$.     
We denote by $(\eta^A_I)_{0\leq A\leq 9;1\leq I\leq 10}$ the local    coordinates on the space  $\mathfrak{p}\otimes T^*_\z\mathcal{P}$
in the basis $(\mathfrak{l}_A\otimes dz^I)_{0\leq A\leq 9;1\leq I\leq 10}$.  Furthermore, by using the splitting $\eta = \nbetao + \nbetaoo$ and the standard representation (see Section \ref{lie-rep} in the Annex) we write  equivalently  $({\nbetao{}^{c}_{d}}_{\mu}    , {\nbetao{}^{c}_{d}}_{j} ,   {\nbetaoo{}^{c}_{\mu}}  ,  {\nbetaoo{}^{c}_{j}}    )$ for the coordinates  on  $   \p\otimes {T}_{\z}^\ast{\PB}$
in the basis $(u^d_{c}    \otimes   dx^{\mu} ,     u^d_{c}    \otimes       \gamma^j , t_c     \otimes  dx^{\mu} , t_c \otimes  \gamma^j      ) $.  The bundle $\mathfrak{p}\otimes T^*\mathcal{P}$ is endowed with local coordinates   $(x^\mu,g,  {\nbetao{}^{c}_{d}}_{\mu}    , {\nbetao{}^{c}_{d}}_{j} ,   {\nbetaoo{}^{c}_{{\mu}}}  ,  {\nbetaoo{}^{c}_{j}} )$. In these coordinates
$\eta$ reads
$ ( \nbetao , \nbetaoo ) =  
    (  {\nbetao{}^{c}_{d}}_{\mu}    {{u}}^{d}_{c}  \otimes  dx^{\mu}  +  {\nbetao{}^{c}_{d}}_{j}    {{u}}^{d}_{c} \otimes \gamma^j         , 
      {\nbetaoo{}^{c}_{\mu}}     {{t}}_{c}   \otimes  dx^{\mu} +   {\nbetaoo{}^{c}_{j}}     {{t}}_{c}   \otimes \gamma^j   ) \in \p\otimes {T}_{\z}^\ast{\PB} 
  $.

We define the following 10-form on $\mathfrak{p}\otimes T^*\mathcal{P}$
(a section of $\Lambda^{10}T^*(\mathfrak{p}\otimes T^*\mathcal{P})$):
\begin{equation}\label{lagrangienPalatiniStart}
{\pmb{\lambda}}   
  := u^{ab}_i \etaun{}_{{ab}}^{(2)}\wedge (d\etazero+\etazero\wedge\etazero)^i\wedge  \etazero{^{(6)}},
\end{equation}
where  
 $\etaun{}_{ab}^{(2)}:=   (  \frac{\partial}{\partial\nbetaoo{}^a}\wedge   \frac{\partial}{\partial \nbetaoo{}^b} ) \iN\nbetaoo{}^{(4)}  = \frac{1}{2}\epsilon_{abcd}\eta^c\wedge \eta^d
 $. Then,   the   WEC action \eqref{WEC-action-total} is   written as  
 $
   \widehat{\mathcal{S}}_{{\hbox{\sffamily{\tiny{WEC}}}}}[\alpha,\omega]   = \int_\mathcal{P}
\varphi^* {\pmb{\lambda}},
$
where $\varphi$ is such that (\ref{identification-forme-section}) holds.

     \subsubsection{First jet bundle}
     
     We introduce the first jet bundle $J^{1}{\cal Z} := J^1(\mathcal{P},\mathfrak{p}\otimes T^*\mathcal{P})$.      A section $\varphi$ of the fiber bundle $\mathfrak{p}\otimes T^*\mathcal{P}$ 
can be seen as a map
$\varphi: \mathcal{P}\longrightarrow \mathfrak{p}\otimes T^*\mathcal{P}$ such that
$\pi_{\mathfrak{p}\otimes T^*\mathcal{P}}\circ \varphi = \hbox{Id}_\mathcal{P}$. 
The first jet space $J^1{\cal Z}$
is the manifold of triplets $(\z,\y,\ypoint)$, where
$(\z,\y)\in \mathfrak{p}\otimes T^*\mathcal{P}$ and
$\ypoint$ is the equivalence class of local sections $\varphi$ of $\mathfrak{p}\otimes T^*\mathcal{P}$
over a neighborhood of $\z$ such that $\varphi(\z) = \y$, for the equivalence relation:
$\varphi_1\simeq \varphi_2$ iff $d(\eta^A_I\circ \varphi_1)_\z = d(\eta^A_I\circ \varphi_2)_\z$,
$\forall I,A$. We then write $[\varphi]_{\z,\y}$ the class of $\varphi$.
Local coordinates on  $J^1(\mathcal{P},\mathfrak{p}\otimes T^*\mathcal{P})$
are $(z^I,\eta^A_I,\eta^A_{I;J})$, where
\[
 \eta^A_{I;J}(\ypoint) = \frac{\partial (\eta^A_I\circ \varphi)}{\partial z^J}(\z)
 \quad\hbox{where}\quad 
 \ypoint = [\varphi]_{\z,\y},
\]
 Equivalently, the first jet space $J^1{\cal Z}$ is identified   (see \cite{helein14})  with $T^\ast\mathcal{P}\otimes_{\p\otimes T^\ast\mathcal{P}} (T(\p\otimes T^\ast\mathcal{P})/T\mathcal{P})$,   the bundle   whose fiber at $({\z},\y)\in \p\otimes  T^\ast\mathcal{P}$ is
the space of linear maps $\ell  :{T}_{\z}\mathcal{P}\longrightarrow T_{(\z,\y)}(\p\otimes  T^\ast\mathcal{P})$
such that  $d(\pi_\mathcal{P})_{(\z,\y)}\circ \ell  = \hbox{Id}_{{T}_{\z}\mathcal{P}}$, which is   canonically identified with
$T^\ast\mathcal{P}_\z\otimes (T_{(\z,\y)}(\p\otimes^\textsc{n} T^\ast\mathcal{P})/{T}_{\z}\mathcal{P})$).

\subsubsection{De Donder--Weyl bundle} 

 Consider the fiber bundle $\Lambda^{10}T^\ast \mathcal{Z}$ of $10$-forms over $\mathcal{Z} := \p \otimes T^{\ast}{\PB}$. By using the
fibration ${\pi}_{\cal Z}:\Lambda^{10}T^\ast\mathcal{\mathcal{Z}} \longrightarrow \mathcal{Z}$ we define a canonical $10$-form $\theta^{(10)}  $ on $\Lambda^{10}T^\ast\mathcal{Z}$ by  $\forall (\z,\y)\in \mathcal{Z}$,
$\forall {\textsf{p}} \in \Lambda^{10}T^\ast_{(\z,\y)}\mathcal{Z}$, $\forall X_1,\cdots, X_{10}\in T_{(\z,\y,{\textsf{p}})}(\Lambda^{10}T^\ast\mathcal{Z})$,
\begin{equation}\label{general-PC}
 \theta^{(10)}_{(\z,\y,{\textsf{p}})}(X_1,\cdots ,X_{10})
:=  {\textsf{p}}( (\pi_\mathcal{Z}){}_*X_1,\cdots, (\pi_\mathcal{Z}){}_*X_{10}) .
\end{equation}
 The $46\ 897\ 636\ 624\ 091$-dimensional  universal Lepage--Dedecker manifold $\Lambda^{10}T^*{\cal Z}$
is far too big.  We   define
the subbundle of  $9$-horizontal forms (following the terminology used in \cite{Cantrijn1,forgergomes})
\[
 \Lambda^{10}_{\mathfrak{1}}T^{\ast}\mathcal{\mathcal{Z}}:=
 \left\{(\z,\y,{\textsf{p}})\in \Lambda^{10}T^*\mathcal{\mathcal{Z}}; \forall v_1,v_2\in {{V}}_{(\z,\y)}{\cal Z},v_1\wedge v_2\iN {\textsf{p}} = 0\right\}.
\]
where   the projection map $\pi_\mathcal{Z}:\mathcal{Z}
\longrightarrow \mathcal{P}$ defines in each tangent space
$T_Z\mathcal{Z}$ a vertical subspace ${{V}}_{(\z,\y)}{\cal Z}:= \hbox{ker}(\pi_\mathcal{Z}{}_{*})$.
We denote by   
${\cal N}_{\hbox{\sffamily{\tiny{DW}}}}  := \Lambda^{10}_{\mathfrak{1}}T^*\mathcal{\mathcal{Z}}$  the {DW}  bundle,   the {{\guillemotleft~multimomentum phase space~\guillemotright}}     of the  
 DW theory. 
   Local coordinates  on   $ {\cal N}_{\hbox{\sffamily{\tiny{DW}}}}  $ are   $(x^{\mu},g, \eta^A_{\mu},\eta^A_j , {\varsigma} ,\psi^{\mu\nu}_A,\psi^{j{\bf \nu}}_A,\psi^{\mu j}_A,\psi^{jk}_A )$, where 
     $(\varsigma,\psi^{\mu\nu}_A,\psi^{j{\nu}}_A,\psi^{{\mu}j}_A,\psi^{jk}_A)$ are the components of
$\psi\in \Lambda^{10}_{\mathfrak{1}}T^\ast_{(\z,\y)}(\p\otimes T^\ast{\PB})$ in the basis
$(\beta^{(4)}\wedge \gamma^{(6)}, d \eta^A_{\mu} \wedge \beta_{\nu}^{(3)}\wedge \gamma^{(6)},
d \eta^A_j\wedge \beta_{\nu}^{(3)}\wedge \gamma^{(6)},   d \eta^A_{\mu}\wedge \beta^{(4)}\wedge \gamma_j^{(5)},
  d \eta^A_j\wedge \beta^{(4)}\wedge \gamma_k^{(5)})$.

  \noindent   The  canonical  $10$-form     \eqref{general-PC}  restricted to ${\cal N}_{\hbox{\sffamily{\tiny{DW}}}}$ is denoted by $\theta^{(10)} \in \Omega^{10}(\Lambda^{10}_{\mathfrak{1}}T^{\ast}{{\cal Z}})$ and  reads
\begin{equation}\label{theta-volterra-general-5}
\begin{array}{cccccc}
{{\theta}}^{(10)} = & \varsigma \beta^{(4)}\wedge \gamma^{(6)}  & + & \psi^{\mu\nu}_Ad \eta^A_{\mu} \wedge \beta_{\nu}^{(3)}\wedge \gamma^{(6)}
 & + & \psi^{j \nu}_Ad\eta^A_j\wedge \beta_{\nu}^{(3)}\wedge \gamma^{(6)} \\
 & & + &  \psi^{\mu j}_Ad\eta^A_{\mu}\wedge \beta^{(4)} \wedge \gamma_j^{(5)}
 & + &  \psi^{jk}_Ad\eta^A_j\wedge \beta^{(4)}\wedge \gamma_k^{(5)}.
 \end{array}  
\end{equation}
   
 \noindent   Since we are interested in normalized sections of $\p\otimes T^\ast\mathcal{P}$, we   actually work on the bundle 
$\Lambda^{10}_{\mathfrak{1}}T^ \ast(\p\otimes^\textsc{n} T^\ast\mathcal{P})$  over 
$\p\otimes^\textsc{n} T^\ast\mathcal{P}$, which is  constructed through a reduction of $\Lambda^{10}_{\mathfrak{1}}T^\ast(\p\otimes T^\ast\mathcal{P})$ (see \cite{helein14}). 
This amounts to impose  
$ \eta^A_j = \delta^i_j $ if $A = i $ and $\eta^A_j  = 0 $ otherwise. We note that, when these constraints are assumed,
the terms with $\psi^{j\nu}_A$ and $\psi^{jk}_A$ vanish and actually don't play any role in the following,
so that     we may forget about coordinates
$(\psi^{j \nu}_A,\psi^{jk}_A)$. Denoting simply by ${{\theta}}^{(10)}$ the restriction to $\Lambda^{10}_{\mathfrak{1}}T^{\ast}(\g\otimes^\textsc{n} T^*\mathcal{P})$ of ${{\theta}}^{(10)}$
given in (\ref{theta-volterra-general-5}),
this leads to the simplification:
\begin{equation}\label{theta-simple}
\theta^{(10)} =   \varsigma \beta^{(4)}\wedge \gamma^{(6)}  + \psi^{\mu\nu}_Ad\eta^A_{\mu} \wedge \beta_{\nu}^{(3)}\wedge \gamma^{(6)}
   +  \psi^{\mu j}_Ad\eta^A_{\mu}\wedge \beta^{(4)}\wedge \gamma_j^{(5)}.
\end{equation} 
     
 \subsection{Legendre transform}\label{LC-1} 
 
         Let $(\z,\y,\ypoint)\in J^1(\mathcal{P},\mathfrak{p}\otimes T^*\mathcal{P})$
and let $\varphi$ be a section such that $[\varphi]_{\z,\y} = \ypoint$. In order to compute
the Legendre transform at $(\z,\y,\ypoint,{\textsf{p}})$ we need to determine the value of the quantity $W(\z,\y,\ypoint,{\textsf{p}})$ which is defined by
$\varphi^*(\theta^{(10)}  - {\pmb{\lambda}}) = W(\z,\y,\ypoint,\pp) \beta^{(4)}\wedge\gamma^{(6)}$, where we denote  $ \beta^{(4)} \wedge \gamma^{(6)} :=  \varphi^* ( \beta^{(4)} \wedge \gamma^{(6)})$ (see \cite{hk2} for details). Using 
the standard representation of   $\p = \g \oplus \mathfrak{t}$,  the  canonical    form \eqref{theta-simple}  is  then given by:    
\begin{equation}\label{theta-volterra-general-56}
\begin{array}{rcccl}
\theta^{(10)}  &= & \varsigma  \beta^{(4)} \wedge \gamma^{(6)}   &  +&   {\po{}^{d}_{c}}{}^{\bf \mu\nu}  d {\nbetao{}^{c}_{d}}_{\mu}   \wedge \beta_{\nu}^{(3)}\wedge \gamma^{(6)}
+    {\po{}^{d}_{c}}{}^{{\mu}j}   d {\nbetao{}^{c}_{d}}_{\mu}  \wedge \beta^{(4)}\wedge \gamma_j^{(5)}    
\\
&& &+ &  
   {\poo{}_{c}}{}^{\bf \mu\nu}   d     {\nbetaoo{}^{c}_{\mu}}        \wedge \beta_{\nu}^{(3)}\wedge \gamma^{(6)}
+  {\poo{}_{c}}{}^{{\mu}j}  d     {\nbetaoo{}^{c}_{\mu}}    \wedge \beta^{(4)}\wedge \gamma_j^{(5)}   . \\
   \end{array}
\end{equation}
The problem we start with
concerns gauge fields on the space-time manifold ${\ST}$ which are    not only   normalized    but    also {\it equivariant} sections, i.e. such that:
\begin{equation}\label{extra-constraint-confII}
   \etao{}^c_{d}{}_{\mu;j}   + [ {\textswab{l}}_j, \nbetao_\mu]{}^c_{d} 
  =   0, 
\quad \quad \quad \quad
   \etaoo{}^c _{\mu;j} +  {\textswab{l}}_i \cdot \nbetaoo{}^c_\mu =    0,    
\end{equation}
To simplify the computation we  choose the right coframe, as we learned from Cartan. Here given some $(\z,\y,\dot{\y},\textsf{p})$,
we replace the arbitrary coframe  $ ( dx^{\mu}, \gamma^{i}  , d  \eta^{A}_{\mu})$ by  $  ( d x^{\mu}, \gamma^{i}  , \delta  \eta^{A}_{\mu})$  in the expression of $\theta{}_{(\z,\y,{\textsf{p}})}^{(10)}$, where  $\delta  \eta^{A}_{\mu}  $   is   given by
  \begin{equation}\label{transformation-coframe-WEC-simple-matrix}
 \delta  \eta^{A}_{\mu}      
=     d \eta^{A}_{\mu}   - \varphi^{\ast} (d \eta^{A}_{\mu} ) = d \eta^{A}_{\mu}   - \eta^{A}_{\mu;\nu}    d x^{\nu}   - \eta^{A}_{\mu;j}  \gamma^j  ,
     \end{equation}
      since    
         $\eta^A_{I;J}(\ypoint)dz^J = d(\eta^A_I\circ \varphi)_\z = (\varphi^*d\eta^A_I)_\z
$. 
Then,  
 \begin{equation}\label{reason-of-the-trick}
 \forall v\in {T}_{\z}\mathcal{P},\quad
\delta  \eta{}^A_{\mu}      (\ell (v))   = 0.
\end{equation}
  We  compute the Legendre correspondence along   equivariant sections,  {i.e.}  which satisfy the   condition \eqref{extra-constraint-confII}. Therefore,  the change of coframe \eqref{transformation-coframe-WEC-simple-matrix}  is given by
  \begin{equation}\label{transformation-coframe-WEC-simple-matrix-01}
  \left\{
\begin{array}{rcl}
  \delta  {\nbetao{}^{c}_{d}}_{\mu}      
     & = &   \displaystyle  
  d   {\nbetao{}^{c}_{d}}_{\mu}   -  \nbetao{}^{c}_{d}{}_{\mu;\nu}    d x^{\nu}   +    [ {\textswab{l}}_j, \nbetao_\mu]{}^{c}_{d}    \gamma^j, 
 \\
     \delta \nbetaoo{}^{c}_{\mu}       & = &   \displaystyle   
      d \nbetaoo{}^{c}_{\mu}    -   \nbetaoo{}^c_{\mu;\nu} d x^{\nu}   +   {\textswab{l}}_j  \nbetaoo{}^{c}_\mu      \gamma^j,    
 \\ 
   \end{array}
\right.
 \nonumber
\end{equation}
Note that, by using the standard representation  of the WEC form \eqref{lagrangienPalatiniStart} (and since
$
 {{\nbetaoo}}{}^{d\mu\nu}_{c} +  {\nbetaoo}{}^{d\nu \mu}_{c} = 0
$),  we obtain
\begin{equation}  \varphi^*{\pmb{\lambda}}       =  -  {{\nbetaoo}}{}_{c}^{d}{}^{\mu\nu}  \left(  \nbetao{}^{c}_{d}{}_{\mu;\nu}  - \frac{1}{2}   [\nbetao_\mu , \nbetao_\nu]^{c}_{d}   \right) \beta^{(4)}\wedge\gamma^{(6)} .  
\end{equation}
 It follows that:
       \begin{equation}\label{prepare-LT2}
   \begin{array}{rcccl}
\displaystyle   {W}  (\z,\y,\dot{\y},{\textsf{p}})     &=&   (\varsigma \circ \varphi )   
   & +&
  \left( {\poo{}_{c}}{}^{\mu\nu} \circ \varphi \right)  \etaoo{}^c_{\mu;\nu}  
 +  \left( {\po{}^{d}_{c}}{}^{\mu\nu}   \circ \varphi   - {{{\nbetaoo}}}{}^{d\mu\nu}_{c}
  \right)  \etao{}^{c}_{d}{}_{\mu;\nu}    
 \\
 && &- &     \left( {\po{}^{d}_{c}}{}^{\mu j}   \circ \varphi \right)  [ {\textswab{l}}_j, \nbetao_\mu]{}^{c}_{d}  -
  \left(    {\poo{}_{c}}{}^{\mu j}  \circ \varphi \right)   {\textswab{l}}_j  \nbetaoo{}_\mu^{c}  
    \\
  && &- &  \displaystyle 
 \frac{1}{2} {{\nbetaoo}}{}^{d\mu\nu}_{c}  [\nbetao_\mu , \nbetao_\nu]^{c}_{d}  .
\end{array}
\end{equation}
 The Legendre correspondence holds on the points with coordinates
$(\z,\y,\ypoint,{\textsf{p}})$ which are critical points of $W$ with
respect to infinitesimal variations of $\ypoint$ which respect the constraints, i.e., such that
$
 {\partial W}/{\partial \nbetao{}^{c}_{d}{}_{\mu;\nu}}  = 0$ 
 and ${\partial W}/{\partial \nbetaoo{}^{c}_{\mu;\nu}}  = 0$. 
The Legendre  condition yields:
\begin{equation} \label{correspondanceLegendre}
  \poo{}_{a}^{\mu\nu}  \circ \varphi = 0 \quad\hbox{and}\quad  \po{}^d_{c}{}^{\mu\nu}  \circ \varphi =  {{\nbetaoo}}{}^{d\mu\nu}_{c} .
\end{equation}
    The   image of the Legendre transform is 
 denoted by
   \begin{equation}\label{constraint}  \mathcal{N}:= \{(\z,\y,{\textsf{p}})\in \Lambda^{10}_{\mathfrak{1}}T^\ast(\p \otimes^\textsc{n} T^ \ast\mathcal{P}) /  \po{}^d_{c}{}^{\mu\nu}    =     {{\nbetaoo}}{}^{d\mu\nu}_{c}  ,     \poo{}_{a}^{\mu\nu}  = 0  \}
   \end{equation}
 Thus, the 
     value of the Hamiltonian function is then the restriction of $W$ at the points where
(\ref{correspondanceLegendre}) holds, i.e. simply:
\begin{equation}\label{valeurHamiltonien}
 H(\z,\y,{\textsf{p}}) =   \varsigma
- \frac{1}{2} {{\nbetaoo}}{}^{d\mu\nu}_{c}  [\nbetao_\mu , \nbetao_\nu]^{c}_{d} -  \left(   {\po{}^{d}_{c}}{}^{\mu j}     [ {\textswab{l}}_j, \nbetao_\mu]{}^{c}_{d}  +  {\poo{}_{c}}{}^{\mu j}    {\textswab{l}}_j  \nbetaoo{}_\mu^{c} \right) .
\nonumber
\end{equation}

\noindent  We   change the coordinates on $\mathcal{N}$ in order to simplify the Hamiltonian function and in such a way that  $\theta^{(10)}$ 
depends on $\eta$ uniquely through the quantity $d \eta + \eta\wedge \eta$.   We set 
$\varkappa{}  :=     \varsigma 
- \frac{1}{2} {{\nbetaoo}}{}^{d\mu\nu}_{c}  [\nbetao_\mu , \nbetao_\nu]^{c}_{d}  -   (   {\po{}^{d}_{c}}{}^{\mu j}     [ {\textswab{l}}_j, \nbetao_\mu]{}^{c}_{d}  +  {\poo{}_{c}}{}^{\mu j}    {\textswab{l}}_j  \nbetaoo{}_\mu^{c}  ) $,
so that  $
  H(\z,\y,{\textsf{p}}) = \varkappa  
$. 
   The canonical  $10$-form $\theta^{(10)}$ on $\mathcal{N}$  then reads  
   \[
   \begin{array}{rcccl}
     \theta^{(10)} &=& \varkappa \beta^{(4)} \wedge \gamma^{(6)}  &+&\displaystyle \left( \frac{1}{2} \poo_{c}{}^{\mu\nu} \beta_{\mu\nu}^{(2)}\wedge\gamma^{(6)} +  \poo_{c}{}^{\mu j} \beta_{\mu}^{(3)} \wedge\gamma_j^{(5)} \right) \wedge   (d \nbetaoo + \nbetao \wedge \nbetaoo )^{c} 
     \\
       && &+&\displaystyle \left(  \frac{1}{2} \po{}^{d}_{c}{}^{\mu\nu} \beta_{\mu\nu}^{(2)}\wedge\gamma^{(6)} +  \po{}^{d}_{c}{}^{\mu j} \beta_{\mu}^{(3)} \wedge\gamma_j^{(5)} \right) \wedge   (d \nbetao + \nbetao \wedge \nbetao )^{c}_{d} . 
     \end{array}
     \]
\noindent 
The
multisymplectic manifold ($\mathcal{N},{{\pmb{\omega}}})$  
has another  construction (see also   \cite{helein14,HeleinVey01a}).  We choose,  as suitable   submanifold of 
  $\Lambda^{10}T^*(\mathfrak{p}\otimes T^*\mathcal{P})$, the total space of the fiber bundle  $ \mathcal{M} := \R\oplus_\mathcal{P}\left(\mathfrak{p}^*\otimes\Lambda^8T^*\mathcal{P}\right)
\oplus_\mathcal{P}\left(\mathfrak{p}\otimes^\textsc{n} T^*\mathcal{P}\right)$ over $\mathcal{P}$. 
 The base $\mathcal{P}$ is equipped with the volume form $\beta^{(4)}\wedge\gamma^{(6)}$ and 
$\varsigma$ is a coordinate on $\R$.
Denote by $(\psi^{\mu\nu},{\psi}^{\mu j},\psi^{jk})$ the $\p^*$-valued coordinates on the fibers of $\p^*\otimes \Lambda^{8}T^*\mathcal{P}$
in the basis $(\beta_{\mu\nu}^{(2)}\wedge \gamma^{(6)}, \beta_{\mu}^{(3)}\wedge \gamma_j^{(5)},\beta^{(4)}\wedge \gamma_{jk}^{(4)})$.
The bundle  $\p^*\otimes \Lambda^{8}T^*\mathcal{P}$ is endowed with the canonical $\p^*$-valued
$8$-form $\psi$    
   defined by (see \cite{HeleinVey01a}): $\forall (\textsf{z},\textsf{p})\in \mathfrak{p}^*\otimes \Lambda^8T^*\mathcal{P}$, $\forall w_1,\cdots,w_8\in T_{(\textsf{z},\textsf{p})}(\mathfrak{p}^*\otimes \Lambda^8T^*\mathcal{P}),$
\[
 \psi_{(\textsf{z},\textsf{p})}(w_1,\cdots,w_8) = \textsf{p}(d\pi_{(\textsf{z},\textsf{p})}(w_1),\cdots,
d\pi_{(\textsf{z},\textsf{p})}(w_8)),
\]
where $\pi = \pi_{\mathfrak{p}^*\otimes \Lambda^8T^*\mathcal{P}}: 
\mathfrak{p}^*\otimes \Lambda^8T^*\mathcal{P}\longrightarrow \mathcal{P}$ is
the canonical projection map. This $\mathfrak{p}^*$-valued 8-form decomposes as
$\psi = \psi_A\mathfrak{l}^A$ where   $\psi_A$ is written  as:
  \begin{equation}
    \psi_A :=   \frac{1}{2} \psi^{\mu\nu}_A \beta_{\mu\nu}^{(2)}\wedge\gamma^{(6)} 
 +  \psi^{\mu j}_A \beta_{\mu}^{(3)} \wedge\gamma_j^{(5)} +  \frac{1}{2}  \psi^{ij}_A \beta^{(4)}  \wedge\gamma_{ij}^{(4)} .
 \end{equation}
\noindent
The final    multisymplectic space  under consideration is    the submanifold    ${\mathcal{N}}_{\circ} := {\cal N} \cap H^{-1}(0)  $ of $\mathcal{M}$ 
which is the intersection of the image of the Legendre correspondence,  defined
by the constraints (\ref{correspondanceLegendre}), with the level set $H^{-1}(0)$. Thus, the canonical  $10$-form  on ${\mathcal{N}}_{\circ}$ has the simple structure: 
$\theta^{(10)}:=    \psi_A\wedge (d\eta+ \eta\wedge \eta )^{A}$.

   \section{The Hamilton  equations}\label{sec:HVDW-equations}

We
consider a point $ (\z,{\textsf{m}},\dot{\textsf{m}}) \in J^{1} ({\PB} , (\p \otimes^\textsc{n} T^ \ast\mathcal{P}) \oplus_{\PB} (\p^* \otimes \Lambda^{8} T^ \ast\mathcal{P}))$, where $\z \in {\PB}$, 
$\m\in (\mathfrak{p}\otimes^\textsc{n}  T^*_\z\mathcal{P})\oplus(\mathfrak{p}^*\otimes^\textsc{n} \Lambda^8T^*_\z\mathcal{P})$ and  $\dot{\m}$ represent the tangent space to a section of the first jet bundle  
$J^1(\mathcal{P},
(\mathfrak{p}\otimes^\textsc{n}  T^*\mathcal{P})\oplus_\mathcal{P}(\mathfrak{p}^*\otimes^\textsc{n} \Lambda^8T^*\mathcal{P}))$
at $(\z,\m)$. Local coordinates on   
$(\mathfrak{p}\otimes^\textsc{n}  T^*\mathcal{P})\oplus_\mathcal{P}(\mathfrak{p}^*\otimes^\textsc{n} \Lambda^8T^*\mathcal{P})$
are $(z^I,\eta^A_\mu,\psi^{\mu\nu}_A,\psi^{\mu j}_A)$,  where  $\eta^A = \eta^A_Idz^I$.
We identify  $\mpoint$ with the equivalence class of sections $\phi$ of  
  $(\mathfrak{p}\otimes^\textsc{n} T^*\mathcal{P})\oplus_\mathcal{P}(\mathfrak{p}^*\otimes^\textsc{n}\Lambda^8T^*\mathcal{P})$
 over  $\mathcal{P}$ such that  $\phi(\z) = \m$,  for the equivalence relation:
\[ 
\phi_1\sim \phi_2
\quad \hbox{iff} \quad
\left\{
\begin{array}{rcl}  d(\eta^A_I\circ \phi_1)(\z) &=& d(\eta^A_I\circ  \phi_2)(\z),\\ 
  d(\psi{_A}{^{\mu\nu}}\circ \phi_1)(\z) &=& d(\psi{_A}{^{\mu\nu}}\circ \phi_2)(\z),\\
    d(\psi{_A}{^{\mu j}}\circ \phi_1)(\z) &=& d(\psi{_A}{^{\mu j}}\circ \phi_2)(\z).\\
   \end{array}    \right.
\]

 The    {HVDW} {equations} in $(\mathcal{N}_\circ, {\pmb{\omega}} )$  consists in a condition on a $10$-dimensional  oriented submanifold ${\pmb{\gamma}}$ of  $(\mathfrak{p}\otimes T^*\mathcal{P})\oplus_\mathcal{P}(\mathfrak{p}^*\otimes\Lambda^8T^*\mathcal{P})$ which says that, for any point
$\textsf{m}$ of coordinates $(z^{I},\eta^A_{\mu},\psi^{\mu\nu}_A,\psi^{\mu j}_A)$ of ${\pmb{\gamma}}$,
if $(X_1,\cdots,X_{10})$ is a basis of the tangent space to ${\pmb{\gamma}}$ at $\textsf{m}$
such that $\beta^{(4)}\wedge\gamma^{(6)}(X_1,\cdots,X_{10}) = 1$, then
\begin{equation}\label{volterra-hamilton-basic}
 X_1\wedge \cdots \wedge X_{10}\iN d\theta^{(10)} = 0  ,
\end{equation}
(see \cite{hk2}). The independence condition $\beta^{(4)}\wedge\gamma^{(6)}(X_1 , \cdots , X_{10}) \neq 0$ means  that such sub-manifolds are locally the graph of some section  $\phi$ of the fiber   
$(\mathfrak{p}\otimes^\textsc{n} T^*\mathcal{P})\oplus_\mathcal{P}(\mathfrak{p}^*\otimes\Lambda^8T^*\mathcal{P})$
over  $\mathcal{P}$.  For any section  $\phi$, we denote by 
$\phi^\circledast d\theta^{(10)}$ the  10-form on   $\mathcal{P}$,   
such that the fiber at  $\z\in \mathcal{P}$ is $T^*_{\phi(\z)}\left[(\mathfrak{p}\otimes^\textsc{n} T^*\mathcal{P})
 \oplus_\mathcal{P}(\mathfrak{p}^*\otimes\Lambda^8T^*\mathcal{P})\right]$, which is defined by: $\forall Z_1,\cdots,Z_{10}\in T_\z\mathcal{P}$,
$\forall v\in T_{\phi(\z)}\left[(\mathfrak{p}\otimes^\textsc{n} T^*\mathcal{P})
 \oplus_\mathcal{P}(\mathfrak{p}^*\otimes\Lambda^8T^*\mathcal{P})\right]$,
\[
 (d\theta^{(10)})_{\phi(\z)}(\phi_*Z_1,\cdots,\phi_*Z_{10},v) = \left((\phi^\circledast d\theta^{(10)})_\z(Z_1,\cdots,Z_{10}),v\right),
\]
or equivalently $
 ((\phi^\circledast d\theta^{(10)})_\z,v) = (-1)^{10}\psi^*(v\iN d\theta^{(10)})$. The   HVDW  equations then  read
$
 \phi^\circledast d\theta^{(10)} = 0
$.
\noindent
 The $11$-plectic form  ${\pmb{\omega}}= d \theta^{(10)}$ on $ {\mathcal{N}}_{\circ} $ is  
 \begin{equation}\label{multisymplectic-form-caseII}
 {\pmb{\omega}} =   d \psi_A \wedge  (d \eta + \eta \wedge \eta)^{A}  +   [d \eta\wedge\eta]^{A} \wedge  \psi_A .
\end{equation}
where we denote $[d \eta\wedge\eta]:= d \eta \wedge \eta - \eta\wedge d \eta $.
Following the same steps as in \cite{helein14},  we evaluate separately the terms in (\ref{multisymplectic-form-caseII}) in view of finding the  Hamilton   equations.  Given some point $\textsf{m}$ of ${\pmb{\gamma}}$ of coordinates $(z^{I},\eta^A_{\mu},\psi^{\mu\nu}_A,\psi^{\mu j}_A)$, we replace
the coframe  
$
    (d z^I, d        \eta{}^{A}_\mu  ,   d        \psi^{\mu\nu}_A   , d        \psi^{\mu j}_A   )
$
at
$\textsf{m}$ by the coframe $
     (d z^I,   \delta    \eta{}^{A}_\mu  ,       \delta       \psi^{\mu\nu}_A  ,   \delta       \psi^{\mu j}_A   )
$, 
where:
\begin{equation}\label{change-of-frame}
\left\{
   \begin{array}{ccl}
    \delta    \eta{}^{A}_\mu    & := &  d      \eta{}^{A}_\mu  
    - \phi^* d  \eta{}^{A}_\mu
\\
          \delta   \psi^{\mu\nu}_A     & := &  d   \psi^{\mu\nu}_A  -    \phi^* d  \psi^{\mu\nu}_A   
   \\
 \delta   \psi^{\mu j}_A     & := &  d   \psi^{\mu j}_A  -    \phi^* d  \psi^{\mu j}_A   
         \end{array}\right.
         \end{equation}
         which is equivalently written as:
         \begin{equation}\label{change-of-frame22}
         \left\{   \begin{array}{ccl}
    d    \eta{}^{A}_\mu    & = &  \delta      \eta{}^{A}_\mu  
  +  \eta{}^A_{\nu;\mu}  d x^{\nu}   +   \eta{}^A_{\mu;j}  \gamma^j
\\
          d   \psi^{\mu\nu}_A     & = &  \delta   \psi^{\mu\nu}_A   
          +
              \psi{}_{A;\rho}^{\mu\nu} d x^{\rho}   +      \psi{}_{A;j}^{\mu\nu}\gamma^j    \\
 d   \psi^{\mu j}_A     & = &  \delta   \psi^{\mu j}_A     +
              \psi{}_{A;\rho}^{\mu j} d x^{\rho}   +      \psi{}_{A;k}^{\mu j}\gamma^k 
         \end{array}
                       \right.
         \end{equation}
     Since    
         $\psi^A_{I;J}(\ypoint)dz^J = d(\eta^A_I\circ \phi)_\z = (\phi^*d\eta^A_I)_\z
$ and  $ \psi^{IJ}_{A;K}   (\ypoint)dz^K = d(\psi^{IJ}_{K}  \circ \phi)_\z = (\phi^*d\psi^{IJ}_{A;K} )_\z$. Note that in the following, we  abuse    notations  $\eta{}^{A}_\mu    :=   \phi^\ast \eta{}^{A}_\mu   $, $\psi^{\mu\nu}_A := \phi^*\psi^{\mu\nu}_A$ and $ \psi^{\mu j}_A  = \phi^* \psi^{\mu j}_A $.

\subsection{Computation of the $11$-plectic form  ${\pmb{\omega}}$ }\label{ole}
Let us denote by ${\pmb{\omega}}_{\mathfrak{1}} := d \psi_A\wedge(d\eta + \eta\wedge\eta)^A$ and $ {\pmb{\omega}}_{\mathfrak{2}} :=   [d \eta\wedge\eta]^A\wedge \psi_A$ such that ${\pmb{\omega}} := 
{\pmb{\omega}}_{\mathfrak{1}} + {\pmb{\omega}}_{\mathfrak{2}}$. 

  
  \subsubsection{The computation of ${\pmb{\omega}}_{\mathfrak{1}} $}
  
  Recall that we work on normalized sections  {i.e.}  on the space  $(\mathfrak{p}\otimes^\textsc{n} T^*\mathcal{P})\oplus_\mathcal{P}(\mathfrak{p}^*\otimes^\textsc{n}\Lambda^8T^*\mathcal{P})$
 over  $\mathcal{P}$.   Then using     the change of coframe \eqref{change-of-frame}, we have 
 \[
\begin{array}{rcl}
 (d \eta{} + \eta{}\wedge\eta{})^{A} &=&  \delta\eta{}_\mu^{A}\wedge d x^{\mu} + 
 \frac{1}{2}\left(\eta{}_{\nu;\mu}^{A}-\eta{}_{\mu;\nu}^{A} + [\eta{}_\mu,\eta{}_\nu]^{A} \right) \beta^{\mu\nu}
\\
&&  \hfill
 -  \left(\eta{}_{\mu;j}^{A} - [\eta{}_\mu,{\textswab{l}}_j]^{A}\right)d x^{\mu} \wedge\gamma^j,
  \end{array}
\]
which in the standard representation is written as:
\[\left\{
\begin{array}{rcl}
(d \eta{} + \eta{}\wedge\eta{})^{a}  & = & \delta\etaoo{}^{d}_\mu\wedge d x^{\mu} +  
 \frac{1}{2}\left(\etaoo{}^{a}_{\nu;\mu}- \etaoo{}^{a}_{\mu;\nu}  + [\etao{}_\mu,\etaoo{}_\nu]{}^{a}{} \right) \beta^{\mu\nu}
\\
& & \hfill - \left(\etaoo{}^{a}_{\mu;j}  - [\etao{}_\mu,{\textswab{l}}_j] {}^{a}{} \right)d x^{\mu} \wedge\gamma^j,
\\
 (d \eta{} + \eta{}\wedge\eta{})_{c}^{d} &=&  \delta\etao{}_{c}^{d}{}_\mu\wedge d x^{\mu} +  
 \frac{1}{2}\left(\etao{}_{c}^{d}{}_{\nu;\mu}- \etao{}_{c}^{d}{}_{\mu;\nu}  + [\etao{}_\mu,\etao{}_\nu]{}_{c}^{d}{} \right) \beta^{\mu\nu}
\\&&  \hfill -
 \left(\etao{}_{c}^{d}{}_{\mu;j}  - [\etao{}_\mu,{\textswab{l}}_j]{}_{c}^{d}{} \right)d x^{\mu} \wedge\gamma^j.
   \end{array}\right.
\]
Note also that $ d\psi_{A}  =  \frac{1}{2} d \psi^{\mu\nu}_{A} \wedge \beta_{\mu\nu}^{(2)}\wedge\gamma^{(6)} +  d\psi^{\mu j}_{A} \wedge \beta_{\mu}^{(3)} \wedge\gamma_j^{(5)} $. Then,    we apply       the change of coframe \eqref{change-of-frame} and  algebraic relations  which are given  in Section \ref{co-yoga} of the Annex:
\[
\begin{array}{rcl}
 d\psi_{A} &=& 
          \frac{1}{2}  \delta   \psi^{\mu\nu}_A   \wedge \beta_{\mu\nu}^{(2)}\wedge\gamma^{(6)} 
          +
                  \frac{1}{2}      \psi{}_{A;\rho}^{\mu\nu} \left( \delta^{\rho}_{\nu}   \beta_{\mu}^{(3)} -  \delta^{\rho}_{\mu}   \beta_{\nu}^{(3)} \right) \wedge\gamma^{(6)} 
              \\
             & &  
             +  \  \delta   \psi^{\mu j}_A     \wedge \beta_{\mu}^{(3)} \wedge\gamma_j^{(5)}
             +
              \psi{}_{A;\mu}^{\mu j}  \beta^{(4)} \wedge\gamma_j^{(5)}
              -      \psi{}_{A;j}^{\mu j} \beta_{\mu}^{(3)} \wedge\gamma^{(6)} .
     \displaystyle
     \end{array}
\]
We translate this expression by using the standard representation and  imposing  the constraints (\ref{correspondanceLegendre}): 
\[  
\left\{
\begin{array}{rcl}
\displaystyle    
 d\poo_{a} & = &  \displaystyle     
    \delta  {\poo{}_{a}}{}^{\mu j}        \wedge {\beta}_{\mu}^{(3)} \wedge \gamma_j^{(5)}     +  
     {\bfpoo{}_{a}}{}_{;\mu}^{\mu j}   {\beta}^{(4)}  \wedge \gamma_j^{(5)}  
  -   {\bfpoo{}_{a}}{}_{;k}^{\mu j}\gamma^k          \wedge {\beta}_{\mu}^{(3)} \wedge \gamma_j^{(5)} , 
 \\
\displaystyle   d\po{}^{d}_{c}  
    & = &   
      \displaystyle            {\varepsilon_{abc}}^{d}   \delta  {\nbetaoo}{}^{a}_\mu   \wedge {d} x^{\mu} \wedge \nbetaoo{}^{b}  \wedge \gamma^{(6)}  +   {\varepsilon_{abc}}^{d}     {\etaoo{}^{a}}_{\mu;\nu}    {\beta}^{\nu\mu} \wedge \nbetaoo{}^{b}  \wedge \gamma^{(6)}      
\\
\displaystyle       &   &       \displaystyle  \hfill +      \      \delta  {\po{}^{d}_{c}}{}^{\mu j}    \wedge {\beta}_{\mu}^{(3)}  \wedge \gamma_j^{(3)}     +   {\bfpo{}^{d}_{c}}{}_{; \mu}^{\mu j}  {\beta}^{(4)} \wedge \gamma_j^{(5)}   
 -   {\bfpo{}^{d}_{c}}{}_{;j}^{\mu j}         {\beta}_{\mu}^{(3)}  \wedge \gamma^{(6)}  . 
 \end{array}
\right.
\]
  Finally, the expression of  $ {\pmb{\omega}}_{\mathfrak{1}} := d \psi  \wedge  (d \eta + \eta \wedge \eta) $, is given by: 
       \begin{equation}\label{dpdeta-caseII}
\begin{array}{ccl}
   {\pmb{\omega}}_{\mathfrak{1}} & = &       {{\epsilon}_{abc}}^{d}   \delta  {\nbetaoo}{}^{a}_\mu \wedge  \etaoo{}^{b}  \wedge     \delta \nbetao{}^{c}_{d}{}_{\sigma} \wedge  \beta^{\mu\sigma}    \wedge    \gamma^{(6)}      
     \\
  &  & 
       +    \left[    \delta  {\po{}^{d}_{c}}{}^{\mu j}    \wedge   \delta \nbetao{}^{c}_{d}{}_{\mu}   +          \delta  {\poo{}_{a}}{}^{\mu j}                \wedge     \delta \nbetao{}^{a}_{\mu}    \right]   \wedge \beta^{(4)} \wedge \gamma_{j}^{(5)}   
    \\
 &  &        +   \ \delta  {\nbetaoo}{}^{a}_\mu   \wedge      \left[ \frac{1}{2}   {{\epsilon}_{abc}}^{d}    \left( \etao{}^{c}_{d}{}_{\tau; \sigma}-\etao{}^{c}_{d}{}_{\sigma;\tau} + [\etao{}_\sigma,\etao{}_\tau]^{c}_{d}{} \right)\beta^{\sigma\tau\mu}    \wedge \nbetaoo{}^{b}  -       {\bfpoo{}_{a}}{}_{;j}^{\mu j}     \beta^{(4)}        \right]   \wedge \gamma^{(6)}       
    \\
 &  &        + \    \delta \nbetao{}^{c}_{d}{}_{\sigma}      \wedge   \left[ \frac{1}{2}   {{\epsilon}_{abc}}^{d}   \left(  {\etaoo{}^{a}_{\mu;\nu}}  -  {\etaoo{}^{a}_{\nu;\mu}} \right)     \beta^{\nu\mu\sigma}       \wedge \nbetaoo{}^{b}       -   {\bfpo{}^{d}_{c}}{}_{;j}^{\mu j}          \beta^{(4)}      \right]    \wedge \gamma^{(6)}       
 \\
  &   &    \displaystyle  +  \   \delta  {\po{}^{d}_{c}}{}^{\mu j}   \wedge   \left( \etao{}^{c}_{d}{}_{\tau;k} - [\etao{}_\tau,{\textswab{l}}_k]^{c}_{d}\right)  \beta^{(4)} \wedge\gamma^{(6)} 
 \\
  &   &    \displaystyle  + \    \delta  {\poo}{}_{a}^{\mu j}        \wedge     \left(\etao{}^a_{\tau;k} - [\etao{}_\tau,{\textswab{l}}_k]^{a}\right) \beta^{(4)} \wedge\gamma^{(6)}  
 .    \end{array}
\end{equation}

  \subsubsection{The computation of ${\pmb{\omega}}_{\mathfrak{2}} $}

Let us compute $ {\pmb{\omega}}_{\mathfrak{2}} :=   [d \eta\wedge\eta]^A\wedge \psi_A$. First, note that  $[d \eta\wedge\eta] ^{A}= [d \eta,\eta_{\mu}]^{A}\wedge dx^{\mu} + [d \eta,{\textswab{l}}_j]^{A}\wedge\gamma^j$, then
\[
 \begin{array}{ccl}
 {\pmb{\omega}}_{\mathfrak{2}} & = & 
  \frac{1}{2}[d \eta,\eta{}_{\nu}]^A\wedge  \psi{}^{\mu\rho}_A(\delta^\nu_\rho\beta_{\mu}^{(3)} - \delta^\nu_\mu\beta_\rho^{(3)})\wedge\gamma^{(6)}
  \\
  & & 
 +  [d \eta,\eta{}_{\nu}]^A\wedge \psi{}^{\mu k}_A\delta^\nu_\mu\beta^{(4)}\wedge \gamma_k^{(5)} -  [d \eta,{\textswab{l}}_j]^A\wedge \psi{}^{\mu k}_A \delta^j_k\beta_{\mu}^{(3)}\wedge \gamma^{(6)},
 \end{array}
\]
so that:
 \begin{equation}\label{deta-phase2}
 \begin{array}{rcl}
 {\pmb{\omega}}_{\mathfrak{2}}  &=&  [d \eta,\eta{}_{\nu}]^{A}\wedge \psi{}^{\mu\nu}_A\beta_{\mu}^{(3)}\wedge\gamma^{(6)}
 +  [d \eta,\eta{}_a]^{A}\wedge \psi{}^{\mu k}_A\beta^{(4)}\wedge\gamma_k^{(5)}
 \\
 &&\hfill  - [d\eta,{\textswab{l}}_j]^{A}\wedge {{\psi}}{}^{\mu j}_A \beta_{\mu}^{(3)}\wedge \gamma^{(6)}.
 \end{array}
\end{equation}
By using the change of coframe given in Equation \eqref{change-of-frame}, we are able to simplify   Equation \eqref{deta-phase2} such that:
\begin{equation}\label{deta-phase4}
 {\pmb{\omega}}_{\mathfrak{2}}  = \left( 
 \psi{}^{\mu\nu}_A[\eta_b,\delta\eta_{\mu}]^A
 +
  {{\psi}}{}^{\mu j}_A[{\textswab{l}}_j,\delta\eta_{\mu}]^A\right) \wedge\beta^{(4)}\wedge\gamma^{(6)}.
\end{equation}
 In  Equation \eqref{deta-phase4} we have a duality product between 
$\psi{}^{\mu\nu}_A$ and $ \left( \hbox{ad}_{\eta_\nu} (\delta\eta_\mu)\right)^{A}  := [\eta_\nu,\delta\eta_{\mu}]^A$, which is equivalently seen as the   product between
$ ( \hbox{ad}_{\eta_{\nu}}^*({{\psi}}^{\mu\nu}) ){}_A$ and $\delta\eta_{\mu}^{A} $, 
where $\hbox{ad}_{\eta_{\nu}}^*$
is the adjoint of $\hbox{ad}_{\eta_{\nu}}$. 
We have also the duality product between  $ {{\psi}}{}^{\mu j}_A$ and $  ( \hbox{ad}_{{\textswab{l}}_j} (\delta\eta_\mu) )^{A}  := [{\textswab{l}}_j,\delta\eta_{\mu}]^A$, which is equivalent with the   product between $ ( \hbox{ad}_{{\textswab{l}_{j}}}^*({{\psi}}^{\mu j})  ){}_{A}    $ and   $\delta\eta_{\mu}^{A} $. We refer to Section \ref{lie-rep} in the Annex for further details on adjoint and coadjoint actions. Hence (\ref{deta-phase4}) reads:

 \begin{equation}\label{deta-final-integrable-representation-01}
  {\pmb{\omega}}_{\mathfrak{2}}  = 
  \left( \left(  \hbox{ad}_{\eta_\nu}^*({{\psi}}^{\mu\nu}) \right){}_{A}  +   \left(  \hbox{ad}_{{\textswab{l}_{j}}}^*({{\psi}}^{\mu j}) \right){}_A   \right)
       \delta \eta{}_{\mu}^{A}       \wedge\beta^{(4)}\wedge\gamma^{(6)}.  
\end{equation}
which, in the standard representation, is equivalently given by  
\begin{equation}\label{deta-final-integrable-representation-01}
          \begin{array}{ccl}
 {\pmb{\omega}}_{\mathfrak{2}}  & = &  
 \left( \left( \hbox{ad}_{\eta_\nu}^*({{\psi}}^{\mu\nu}) \right){}^{d}_{c} 
 +  \left( \hbox{ad}_{{\textswab{l}_{j}}}^*({{\psi}}^{\mu j})   \right){}^{d}_{c} \right)
       \delta \nbetao{}^{c}_{d}{}_{\mu}       \wedge \beta^{(4)}\wedge\gamma^{(6)} 
         \\
  &   & 
  +  \left(\left(  \hbox{ad}_{\eta_{\nu}}^*({{\psi}}^{\mu\nu}) \right){}_a + \left( \hbox{ad}_{{\textswab{l}_{j}}}^*({{\psi}}^{\mu j})  \right){}_{a}\right)
    \delta  {\nbetaoo}{}^{a}_\mu  \wedge \beta^{(4)}\wedge\gamma^{(6)}. 
  \\
  \end{array}
\end{equation}
    Working on the submanifold of  constraints ${\cal N}$, see   \eqref{correspondanceLegendre}, and using Lemma \ref{lem:constcalcul1}     (see Section \ref{Einstein-Spin}),  Equation  \eqref{deta-final-integrable-representation-01} yields:
 \begin{equation}\label{deta-final-integrable-representation-07}
 \begin{array}{rcl}
  {\pmb{\omega}}_{\mathfrak{2}}    &=&    
    \delta \nbetao{}^{c}_{d}{}_{\mu}     \wedge \left(d x^{\mu} \wedge      {{\epsilon}_{abc}}^{d}  {\etao{}^{a}_{a'}} \wedge  {\etaoo{}^{a'}}     \wedge   {\etaoo{}^{b}}    + 
          [  {\bfpo}{}^{\mu j}   , {{\textswab{l}_{j}}}       ]{}^{d}_c          \beta^{(4)} \right) \wedge\gamma^{(6)}    
 \\
  & &          + \  \delta  {\nbetaoo}{}^{a}_\mu  \wedge     ( {\bfpoo}{}^{\mu j}    {{\textswab{l}_{j}}}    )_a \beta^{(4)}\wedge\gamma^{(6)}. 
    \end{array}
\end{equation}

\subsection{HVDW--WEC equations}\label{WEC--HVDW-equations}
Collecting  (\ref{dpdeta-caseII}) and (\ref{deta-final-integrable-representation-07}), the $11$-plectic form $ {\pmb{\omega}}  := d \theta^{(10)} $ is then written as:
  \begin{equation}\label{omega-prepare-2}
\begin{array}{ccl}
 {\pmb{\omega}}  & = &  \ \hbox{\small non linear terms in} \ \delta  {\eta}{}^{A}_\mu \ \hbox{\small and} \  \delta  \psi{}_{A}^{\mu j}  
 \\
 &  &       + \   \delta  {\nbetaoo}{}^{a}_\sigma   \wedge      \left[    2 {{\Upsilon}}_a  
      \wedge dx^{\mu}    
    -    \left(    {\bfpoo{}_{a}}{}_{;j}^{\sigma j}  -      {\bfpoo{}_{b}}{}^{\sigma j}    {{\textswab{l}_{j}}{}^{b}_{a}}    
\right)     
 \beta^{(4)} 
  \right]  
    \wedge \gamma^{(6)}      
                  \\
 &  &       +  \   \delta \nbetao{}^{c}_{d}{}_{\sigma}      \wedge      \left[    2{{\Sigma}}_c{}^{d} 
     \wedge dx^{\mu}
   -  \left(  {\bfpo{}^{d}_{c}}{}_{;j}^{\sigma j}   -   [  {\bfpo}{}^{\sigma j}   , {{\textswab{l}_{j}}}       ]{}^{d}_c         \right)           \beta^{(4)}    \right]   \wedge \gamma^{(6)}       
  \\
  &   &      +  \   \delta  {\po{}^{d}_{c}}{}^{\mu j}   \wedge  \left(\etao{}^{c}_{d}{}_{\mu;k} +  [ {\textswab{l}}_k, \etao{}_\mu ]^{c}_{d}   \right) \beta^{(4)}  \wedge\gamma^{(6)}   
    \\
  &   &      +   \   \delta  {\poo{}_{c}}{}^{\mu j}        \wedge     \left(\etaun{}^{c}_{\mu;k} +  \etaun{}_\mu^{c}  {\textswab{l}}_k  \right) \beta^{(4)}  \wedge\gamma^{(6)}  
. \\
 \end{array}
 \end{equation}
We have recognized   the Einstein $3$-forms ${{\Upsilon}}_a:=   \frac{1}{2}  {{\epsilon}_{abc}}^{d}   (     d \etao{}^{d}_{c}{} + \etao{}^{d}_{c'}  \wedge \etao{}^{c'}_{c}       )  \wedge  \nbetaoo{}^{b}   $ and the Spin $3$-forms ${{\Sigma}}_c{}^{d}  :=  \frac{1}{2}    {{\epsilon}_{abc}}^{d}    (     d \etaoo{}^{a}{} + \etao{}^{a}_{a'}  \wedge \etao{}^{a'}       )   \wedge  \nbetaoo{}^{b}   $, which are given by Definitions  \eqref{Einstein-3-form} and    \eqref{Spin-forms}, respectively   (see  Section \ref{Einstein-Spin} in the Annex).     Since  
${{\Upsilon}}_a \wedge   dx^{\rho}  =    \frac{1}{3!}     {\epsilon}^{\rho\lambda\mu\nu}     {{\Upsilon}}{}_{a}{}_{\lambda\mu\nu}  \beta^{(4)} $ and ${{\Sigma}}{}_{c}{}^{d}   \wedge dx^{\rho}    = \frac{1}{3!}       {\epsilon}^{\rho\lambda\mu\nu}   {{\Sigma}}{}_{c}{}^{d}{}_{\lambda\mu\nu} \beta^{(4)}$, Equation  
(\ref{volterra-hamilton-basic}) yields:  
\begin{equation}\label{volterra-hamilton-synt-WEC-01}
 \begin{array}{rcl}
X_1\wedge \cdots \wedge X_{10}\iN d\theta^{(10)}  
  &  = &       
 +    \       \delta  {\nbetaoo}{}^{a}_\mu      \left(  \left(   \frac{2}{3!} \right)
   {\epsilon}^{\sigma\lambda\mu\nu}     {{\Upsilon}}{}_{a}{}_{\lambda\mu\nu}    -       {\bfpoo{}_{a}}{}_{;j}^{\sigma j}  +      {\bfpoo{}_{b}}{}^{\sigma j}    {{\textswab{l}_{j}}{}^{b}_{a}}    
    \right)           
   \\
  &  &         +  \     \delta \nbetao{}^{c}_{d}{}_{\sigma}     \left(   \left(   \frac{2}{3!} \right)
 {\epsilon}^{\sigma\lambda\mu\nu}   {{\Sigma}}{}_{c}{}^{d}{}_{\lambda\mu\nu}   -   {\bfpo{}^{d}_{c}}{}_{;j}^{\sigma j}   +   [  {\bfpo}{}^{\sigma j}   , {{\textswab{l}_{j}}}       ]{}^{d}_c                   \right)  
  \\
  & &\displaystyle +  \  \delta  {\po{}^{d}_{c}}{}^{\mu j}      \left(\etao{}^{c}_{d}{}_{\mu;k} +  [ {\textswab{l}}_k, \etao{}_\mu ]^{c}_{d}  \right)  
\\
&&\displaystyle  
+   \    \delta  {\poo{}_{c}}{}^{\mu j}       \left( \etao{}^{c}_{\mu;k} +  [ {\textswab{l}}_k, \etao{}_\mu ]^{c}  \right) , 
  \end{array}
\end{equation}
where   the first   line  in the right hand side  of (\ref{omega-prepare-2}) do  not contribute because of 
terms quadratic in $\delta(\cdot)$. 
      \begin{prop}\label{HVDW-001}{\em
The HVDW--WEC equations $\phi^\circledast d\theta^{(10)} = 0$ (see \eqref{volterra-hamilton-basic})
yields   the following   system of equations:
 \begin{equation}
\begin{tikzpicture}
    \matrix (m) [matrix of nodes,column 1/.style={anchor=west}]
   {
  $   \left( \frac{2}{3!} \right) {{\epsilon}}^{\sigma\lambda\mu\nu}    {{\Sigma}}{}_{a}{}^{b}{}_{\lambda\mu\nu}     =   \Xio{}_{a}{}^{b}{}^{\sigma} $    \\
  $  \hspace{0.2cm}  \left( \frac{2}{3!} \right)  {{\epsilon}}^{\sigma\lambda\mu\nu}  {{\Upsilon}}{}_{a}{}_{\lambda\mu\nu}  = \Xioo{}_{a}{}^{\sigma} $     \\
  $\hspace{0.25cm} \hspace{0.3cm} \etao{}_{\mu;k} +  [ {\textswab{l}}_k, \etao{}_\mu ] = 0$     \\
   $\hspace{0.25cm}  \hspace{0.74cm}  \etaoo{}_{\mu;k} +    {\textswab{l}}_k  \etaoo{}_\mu   = 0$  \\
     };
       \draw [decoration={brace,amplitude=0.5em},decorate,ultra thick,gray]
        (m-4-1.south -| m.west) -- node[black, left=0.7em]   {$\textsf{{\small {{HVDW--WEC}}}}$} (m-1-1.north -| m.west);
        
    \draw [decoration={brace,amplitude=0.5em},decorate,ultra thick,gray]
        (m-1-1.north -| m.east) -- node[black, right=0.7em]   {$\textsf{{\small {{Einstein--Cartan}}}}$} (m-2-1.south -| m.east);
           \draw [decoration={brace,amplitude=0.5em},decorate,ultra thick,gray]
        (m-3-1.north -| m.east) -- node[black, right=0.7em] {${\textsf{\small {{Equivariance}}}}$} (m-4-1.south -| m.east);
\end{tikzpicture}
\nonumber
\end{equation}
 \noindent where  we   use the   notations
$    \Xio{}_{c}^{d}{}^{\mu} : =     
    {\bfpo{}^{d}_{c}}{}_{;j}{}^{\mu j}  + [ {\textswab{l}}_{j} ,      {\bfpo}{}^{\mu j} ]{}^{d}_{c}  $    
   and $    \Xioo{}_{a}{}^{\mu} :  =   
             {\bfpoo{}_{a}}{}_{;j}^{\mu j}        -      {\bfpoo{}_{b}}{}^{\mu j}    {{\textswab{l}_{j}}{}^{b}_{a}}$.  
}
\end{prop}
   
  \noindent The Hamilton  equations are   composed of the Einstein--Cartan system of equations together with the equivariance condition for the for the 1-form $\eta$. Note that the latter   is not assumed a priori but is obtained
by {{unfolding the dynamics}}:   the   fields $     {\bfpo{}^{d}_{c}}{}^{\mu j}   $ and $    {\bfpoo{}_{c}}{}^{\mu j} $  plays the role of  Lagrange multipliers
for the     constraints given by Equation \eqref{extra-constraint-confII}.  

   \begin{prop}\label{HVDW-002}
      {\em  We denote ${\overset{\mathfrak{0}}{p}{}^{d}_{c}}{}^{\mu j}  := \displaystyle   {g}^{d}_{d'}    {\bfpo{}^{d'}_{c'}}{}^{\mu j}      ({g}^{-1}){}^{c'}_c $ and $      \overset{\mathfrak{1}}{{{p}}}{}^{\mu j}_{a}  :=   {\bfpoo{}_{a'}}{}^{\mu j}      ({g}{}^{-1})^{a'}_{a}  $.   The HVDW--WEC equations $\phi^\circledast d\theta^{(10)} = 0$ 
 yields  the  following Einstein--Cartan system of equations:
   \begin{equation}\label{ECsystem2} 
 \left\{
 \begin{array}{lcl}
   {G}{^b}_a & = & \frac{1}{2} \rho_j\cdot p{_a}{^{bj}}\\
    {T}{^a}_{cd} & = & - \left( \textsf{h}_{de}\delta^a_{a'}\delta^{c'}_c
 + \frac{1}{2}\delta^{c'}_{a'}(\delta^a_d\textsf{h}_{ce}
 - \delta^a_c\textsf{h}_{de})\right) \rho_j\cdot p{_{c'}}{^{ea'j}} , 
 \end{array}
 \right.
\end{equation}
where  $p{_a}{^{bcj}} := \momo{_a}{^{b\sigma j}} e_{\sigma}^{c}$ and $p{_a}{^{bj}} := \momun{_a}{^{\sigma j}} e_{\sigma}^{b}$.
}
 \end{prop}

\noindent \Proof. The equivariance   condition found in Proposition \ref{HVDW-001} 
 is equivalent to say that there exists 
  $\g$-valued functions $A_{\mu} (x) $ and $\mathfrak{t}$-valued functions $e_\mu(x)$,  which depends only on $x \in {\cal X}$   such that (see  Equation \eqref{in-a-trivialization}) $\forall x\in {\cal X},\forall g\in \G$,
\begin{equation}\label{etao-etaoo-eq}
 \etao{}^{a}_{b}{}_\mu(x,g) = (g^{-1})^{a}_{b'} {A}_{\mu a'}^{b'}(x)g^{a'}_b, \quad\quad \etaoo{}^{a}_{\mu} (x,g) = (g^{-1})^{a}_{a'} e_{\mu}^{a'} (x).
\nonumber\end{equation}
Using   Lemmata \ref{lem:algebraic-einstein} and   \ref{lem:algebraic-spin} of the Annex,     ${{\Upsilon}}_a :=  {{G}}_{a'} g^{a'}{}_{a}$ and ${{\Sigma}}{}_{c}{}^{d} :=  {{H}}_{c'}{}^{d'} g^{c'}_{c} (g^{-1})_{d'}^{d} $, respectively.   
\noindent

\begin{equation}\label{soluWEC-HVDW}
 \left\{ \begin{array}{rcl}
  \left(   \frac{1}{3!} \right)
  {{\epsilon}}^{\sigma\lambda\mu\nu}  {{G}}{}_{a}{}_{\lambda\mu\nu}   & = &  \frac{1}{2}   \overset{\mathfrak{1}}{p}{}_{a}{}^{\sigma j}_{;j}        
\\
  \left(   \frac{1}{3!} \right)
     {{\epsilon}}^{\sigma\lambda\mu\nu}    {{H}}{}_{a}{}^{b}{}_{\lambda\mu\nu}     & = &       \frac{1}{2}     {\overset{\mathfrak{0}}{{{p}}}{}_{a}{}^{b}}{}^{\sigma j}_{;j}  ,        \end{array}
 \right.
\end{equation}

   \noindent Then, we use Lemmata   \ref{Einstein-002} and   \ref{Spin-002} so that  the system of Equations  \eqref{soluWEC-HVDW} is equivalent to:
   \begin{equation}\label{soluWEC-HVDW2}
 \left\{ \begin{array}{rcl}
 \displaystyle     G^{b}{_{a}}  e_{b}^{\sigma}   & = &   
 \frac{1}{2}   \overset{\mathfrak{1}}{p}_{a}{}^{\sigma j}_{;j}         
\\
  \displaystyle    {\textsf{h}}^{bb'}     \left(    T^{c}{}_{b'c}   
    e_{a}^{\sigma}       
     +
   T^{c}{}_{ca}    e_{b'}^{\sigma}        
  +  T^{c}{}_{ab'}     e_{c}^{\sigma}      \right)    & = &   \displaystyle      {\overset{\mathfrak{0}}{{{p}}}{}_{a}{}^{b}}{}^{\sigma j}_{;j}   .     \end{array}
 \right.
\end{equation}
The first line of the system \eqref{soluWEC-HVDW2} yields  $G^{b}{_{a}}  e_{b}^{\sigma}    e^{b'}_{\sigma}     =  \frac{1}{2}  \overset{\mathfrak{1}}{p}_{a}{}^{\sigma j}_{;j}  e^{b'}_{\sigma}  \Rightarrow  G^{b'}{_{a}}  =   \frac{1}{2} \overset{\mathfrak{1}}{p}_{a}{}^{\sigma j}_{;j}  e^{b'}_{\sigma} =   \frac{1}{2}   {p}{}_{a}{}^{b' j}_{;j}  $. 
   Analogously, $ {\overset{\mathfrak{0}}{{{p}}}{}_{a}{}^{b}}{}^{\sigma j}_{;j}     = {\overset{\mathfrak{0}}{{{p}}}{}_{a}{}^{b}}{}^{g j}_{;j}   e_{g}^{\sigma}$. Then, the second line in the system of equations \eqref{soluWEC-HVDW2} yields:  
        \begin{equation}\label{ECsystem3} 
 {\overset{\mathfrak{0}}{{{p}}}{}_{a}{}^{b}}{}^{c j}{}_{;j}   =   
   {\textsf{h}}^{bb'}     \left(    T^{d}{}_{b'd}   
       \delta_a^{c}       +    T^{d}{}_{da}          \delta_{b'}^{c}  
  +  T^{d}{}_{ab'}     \delta_d^{c}         \right)   
 =   
  {\textsf{h}}^{be}      T^{d}{}_{ed}   
       \delta_a^{c}     -   {\textsf{h}}^{bc}      T^{d}{}_{ad}       +  {\textsf{h}}^{bd}  T^{c}{}_{ad} , 
 \nonumber
\end{equation}
which is equivalent  to $\hbox{T}{^a}_{cd} = - \left( \textsf{h}_{de}\delta^a_{a'}\delta^{c'}_c
 + \frac{1}{2}\delta^{c'}_{a'}(\delta^a_d\textsf{h}_{ce} - \delta^a_c\textsf{h}_{de})\right)
 p{_{c'}}{^{ea'j}}_{;j}$ (see \cite{HeleinVey01a} for further details).
        \hfill $\square$

Since the Lorentz group 
  $SO(3,1)$ is not compact we cannot conclude that the right hand side  of the HVDW--WEC equations (\ref{soluWEC-HVDW}) vanish in general (as opposed to the case of the Yang--Mills system, see \cite{helein14}).   One way to overcome this difficulty, (see also    \cite{HeleinVey01a}),  is to suppose that the field $\phi^*\psi_{a}{}^{\mu j}$ and $\phi^*\psi_{a}{}^{b}{}^{\mu j}$ have compact support in ${\cal P}$ or decay at infinity. Then, the right hand side  of (\ref{ECsystem2}) vanish and the system of equations  reduces to the Einstein--Cartan system  in vacuum.
\\
\\
\noindent {\textsf{Conclusion}}   --- 
In the context of {\it{liquid fiber  bundles}},  it   is    difficult to give any  meaning to a pre-determined  space--time   topology. 
Nonetheless, 
the     theory  which is  presented  in this       paper (as in Toller's approach \cite{Toller1978})  allows     to prescribe   the  topology of the space--time manifold ({e.g.}       globally hyperbolic space-times, where    the topology  is given by  some  foliation of   Cauchy hypersurfaces). 
 In addition, the $10$-plectic formulation of WEC gravity    might shed new  light on    cosmology and   could open a road to the question of dark energy.  In particular, the point to address is     to see whether the right hand side  of Equations 
(\ref{soluWEC-HVDW}) could be interpreted as a dark source.
 This  would  entails an  interpretation of the multimomenta   $ {p}_{a}{}^{b}{}^{c j}    $ and $ {p}_{a}{}^{c j}   $ and   of  the   relevant  hypotheses to consider  on those fields  which would  implicate that the right hand side  of
(\ref{soluWEC-HVDW})   vanish or, within  the  perspective of the cosmological constant,  is independent of the point of the space-time manifold. 
 

\section{Annex}

  \subsection{Lie algebra and representations}\label{lie-rep}

\noindent   We denote by    $\mathcal{R}:\mathfrak{G}\rightarrow  GL(\vec{\mathbb{M}})$   the standard representations of $\G$. 
We fix $ \mathcal{R}(g)({E}_b) = {E}_a g{^a}_b$,  $\forall g\in \mathfrak{G}$, $\forall 0\leq a,b\leq 3$, where $g{^a}_b$ are  the coefficients of  
 $\mathcal{R}(g)$ in the basis $({E}_a)_{0\leq a \leq 3}$. We      denote  also     $ {\mathcal{R}}: {\g}\rightarrow {\bf gl}(\vec{\mathbb{M}})$  the standard representations of  the  Lie algebra $\g$ of $\G$. Analogously,    $\forall 0\leq a,b\leq 3$, $ {{{\mathcal{R}}}}(\xi)({E}_b) = {E}_a \xi{^a}_b$, where 
  $\xi{^a}_b$ are the coefficients 
of  ${{\mathcal{R}}}(\xi)$ in the basis $({E}_a)_{0\leq a \leq 3}$.   Note that  $\xi^{ab}+\xi^{ba}=0$, where
$\xi^{ab} = \xi{^a}_{b'}\textsf{h}^{b'b}$. In addition,  $\forall 1\leq i\leq 6,\  {\textswab{l}}_i$   is  identified with the matrix with coefficients $u^a_{ib}$; we also note $u^{ab}_i:= u^a_{ib'}\textsf{h}^{b'b}$ and   $u^{ab}_i+u^{ba}_i=0$. Then,  $ (t_a ,  u^a_{ib})
  :=  (t_0 , \cdots t_3 , u^a_{4b} ,  \cdots , u^a_{9b} ) $  is  a basis of $ {\p} $   whereas 
 $(t^{a},u^{ib}_a)    :=  (t^0 , \cdots t^3 , u_a^{4b} ,  \cdots , u_a^{9b} ) $ is a basis  of    $ {\p}^*$. Finally, we  consider  the vector subspace 
$\vec{\Bbb{M}}\wedge \vec{\Bbb{M}}^*:= \{ t{^a}_bE{_a}^b\in \vec{\Bbb{M}}\otimes\vec{\Bbb{M}}^*; t{^a}_{b'}\textsf{h}^{b'b} + t{^b}_{a'}\textsf{h}^{a'a} = 0\}$,  where   $E{_a}^b:= E_a\otimes E^b$.   The standard representation $\mathcal{R}$ of $\mathfrak{G}$ induces the map
$\mathfrak{G}\longrightarrow \vec{\mathbb{M}}\wedge \vec{\mathbb{M}}^*$, $g\longmapsto g{^a}_bE{_a}^b$. 
 \\
\\
\noindent  {\textsf{Adjoint action}} ---  The restriction to $\mathfrak{G}$ of the adjoint representation of $\mathfrak{P}$ on $\mathfrak{p}$
reads $\forall \xi\in \mathfrak{p},$
\[
 \hbox{Ad}_g(\xi{^a}_bE{_a}^b,\xi^aE{_a}) = \left((g{^a}_{a'}\xi{^{a'}}_{b'}(g^{-1}){^{b'}}_b)E{_a}^b,\,g{^a}_{a'} \xi^{a'}E_a\right).
\]
\noindent  {\textsf{Coadjoint action}} --- The coadjoint action of $\mathfrak{G}$ on $\mathfrak{p}^*$ is defined by:
$\forall g\in \mathfrak{G}$, $\forall \lambda\in \mathfrak{p}^*$, $\hbox{Ad}_g^*\lambda$
is the vector in $\mathfrak{p}^*$
such that:
$
\forall \xi\in \mathfrak{p}, 
(\hbox{Ad}_g^*\lambda)(\xi):= \lambda(\hbox{Ad}_g\xi).
$  Then 
\[
 \begin{array}{ccl}
  (\hbox{Ad}_g^*\lambda)(\xi)  & = & \frac{1}{2}\left(g{^{a'}}_{a}\lambda{_{a'}}^{b'}(g^{-1}){^{b}}_{b'}\right)\xi{^{a}}_{b}
+ \left(g{^{a'}}_{a}\lambda_{a'}\right)\xi^{a}.
 \end{array}
\]
 At the level of Lie algebra,  the coadjoint action of $\mathfrak{p}$ on $\mathfrak{p}^*$ is defined by:
$\forall \xi\in \mathfrak{p}$, $\forall \lambda\in \mathfrak{p}^*$, $\hbox{ad}_\xi^*\lambda$
is the vector in $\mathfrak{p}^*$
such that:
$
\forall \zeta\in \mathfrak{p}, 
(\hbox{ad}_\xi^*\lambda)(\zeta):= \lambda(\hbox{ad}_\xi\zeta) = \lambda([\xi,\zeta])
$. 
This gives us:
\begin{equation}\label{coadjoint-dual}
   (\hbox{ad}_\xi^*\lambda)(\zeta)  = \frac{1}{2}\left(\xi{^{c}}_{a} \lambda{_{c}}^{b} - \lambda{_a}^c\xi{^{b}}_{c}
- 2\lambda_a\xi{^{b}}\right)\zeta{^{a}}_{b} + \left(\xi{^{a}}_{b}\lambda_a\right)\zeta{^{b}}
 \end{equation}

\subsection{Coframe yoga}\label{co-yoga}

For any form ${{\alpha}} \in \Omega^{\ast} ({\ST})$, for any 
multivector field $v := v_1 \wedge \cdots \wedge v_p \in \mathfrak{X}^{p} ({\ST})$, we have $v  \iN {{\alpha}}  = (v_1 \wedge \cdots \wedge v_p) \iN {{\alpha}} := v_{p} \iN \cdots \iN v_1 \iN {{\alpha}}  $. We consider a moving frame $
 \left(\partial_\mu,\rho_{i}\right) := \left(\partial_0 , \cdots , \partial_3 , \rho_1 , \cdots , \rho_6 \right)$  and its dual moving coframe $\left(dx^\mu,\gamma^{i}\right) := \left(dx^0 , \cdots , dx^3 , \gamma^{1} , \cdots , \gamma^{6} \right)$ defined on the total space of the  pseudo-orthonormal  frame bundle (see Section \ref{subsubsec:lift-principal-bundle}).
 Consider  the volume  $4$-form  $\beta^{(4)} := dx^{0} \wedge  \cdots \wedge dx^{3} $.  We define       the family   of   basis   $p$-forms $\beta_{\mu_{1} \cdots \mu_{4-p} }^{(p)}$  such that:
\begin{equation}
 \begin{array}{rccccl}
\beta^{(3)}_{\mu} &:=&  
\frac{\partial}{\partial x^{\mu}}  \iN \beta^{(4)};
&  
 \beta^{(2)}_{\mu\nu} :=
\frac{\partial}{\partial x^{\nu}}   \iN \beta^{(3)}_{\mu};
&  
\beta^{(1)}_{\mu\nu\rho} :=
\frac{\partial}{\partial x^{\rho}}   \iN \beta^{(2)}_{\mu\nu};
&  
\beta^{(0)}_{\mu\nu\rho\sigma} :=
\frac{\partial}{\partial x^{\sigma}}   \iN \beta^{(1)}_{\mu\nu\rho}.
\nonumber
 \end{array}
\end{equation}
 \noindent
and we denote by $\beta^{\mu_1 \cdots \mu_p} := dx^{\mu_1} \wedge \cdots \wedge dx^{\mu_p}$.   We have $\beta^{\sigma} \wedge  \beta_\mu^{(3)}  = \delta^{\sigma}_\mu   \beta^{(4)}$ and   the   following algebraic  relations (see also \cite{HeleinVey01a}): 
 \begin{equation}
\left\{
\begin{array}{rcl}
\beta^{\sigma} \wedge  \beta_{\mu\nu}^{(2)}  &= &  \delta^{\sigma}_\nu \beta_\mu^{(3)}  -    \delta^{\sigma}_\mu \beta_\nu^{(3)} , 
\\
 \beta^{\mu\nu} \wedge {{\beta}}_{\rho\sigma}^{(2)}   &=& \delta^{\mu\nu}_{\rho\sigma}
  \beta^{(4)}   ,  
  \\
\beta^{\sigma} \wedge   \beta^{(1)}_{\mu\nu\rho}  &=&  \delta^{\sigma}_\mu \beta_{\nu\rho}^{(2)} + \delta^{\sigma}_\nu \beta_{\rho\mu}^{(2)}  + \delta^{\sigma}_\rho \beta_{\mu\nu}^{(2)}  ,
\\
\beta^{\sigma\kappa} \wedge  \beta_{\mu\nu\rho}^{(1)}  &= &  \delta^{\sigma\kappa}_{\nu\rho} \beta_\mu^{(3)}  
+ \delta^{\sigma\kappa}_{\nu\rho} \beta_\mu^{(3)}  
+ \delta^{\sigma\kappa}_{\nu\rho} \beta_\mu^{(3)}  ,  \\
 \end{array}
\right.
\nonumber
\end{equation}
where $\delta^{\mu\nu}_{\rho\sigma} :=  (\delta^\mu_\rho\delta^\nu_\sigma 
- \delta^\mu_\sigma\delta^\nu_\rho)$.   Analogously, we consider   the volume $4$-form  $e^{(4)} = e^{0} \wedge  \cdots \wedge e^{3}$ and        define   the family   of   basis   $p$-forms  $e_{a_{1} \cdots a_{(4-p)} }^{(p)}$ such that:
\begin{equation}
 \begin{array}{rccccl}
 e^{(3)}_{a} &:=&  
\frac{\partial}{\partial e^{a}} \iN e^{(4)};
&  
 e^{(2)}_{ab} := \frac{\partial}{\partial e^{b}}   \iN e^{(3)}_{a};
&
e^{(1)}_{abc} :=
\frac{\partial}{\partial e^{c}}  \iN e^{(2)}_{ab};
&
e^{(0)}_{abcd} :=
\frac{\partial}{\partial e^{d}}  \iN e^{(1)}_{abc}.
\nonumber
\end{array}
\end{equation}
and  $e^{a_1 \cdots a_p }:= e^{a_1} \wedge \cdots e^{a_p}$. We have $\beta^{\sigma} \wedge  \beta_\mu^{(3)}  = \delta^{\sigma}_\mu   \beta^{(4)}$, $e^{g} \wedge  e_a^{(3)} =  \delta^{g}_a   e^{(4)}$ and   the   following algebraic  relations (see also \cite{HeleinVey01a}): 
\begin{equation}
  \left\{
\begin{array}{rcl}
 e^{g} \wedge  e_{ab}^{(2)}  &= &  \delta^{g}_b e_a^{(3)}  -    \delta^{g}_a e_b^{(3)}  
,   \\
e^{gh} \wedge  e_{ab}^{(2)}  &=&  \delta^{gh}_{ab}   e^{(4)}, 
\\
e^{g} \wedge   e^{(1)}_{abc}  &=&  \delta^{g}_a e_{bc}^{(2)} + \delta^{g}_b e_{ca}^{(2)}  + \delta^{g}_c e_{ab}^{(2)}  , 
\\
e^{gh} \wedge  e_{abc}^{(1)}  &= &\delta^{gh}_{bc} e^{(3)}_{a} +
\delta^{gh}_{ca} e^{(3)}_{b} +
\delta^{gh}_{ab} e^{(3)}_{c}.
  \end{array}
\right.
\nonumber
\end{equation}
   Finally, we  denote $\gamma^{(6)} := \gamma^1\wedge \cdots \wedge \gamma^6$,  
 $\gamma^{i_1 \cdots i_q} := \gamma^{i_1} \wedge \cdots  \wedge\gamma^{i_q}$ and we   define   the family   of   basis   $q$-forms  $\gamma_{i_{1} \cdots i_{(6-q)} }^{(q)}$ such that:
\begin{equation}
 \begin{array}{rccccl}
\gamma^{(5)}_{i} &:=& \rho_i \iN \gamma^{(6)};
&  
 \gamma^{(4)}_{ij} :=
  \rho_{j} \iN \gamma^{(5)}_{i};
&  
\gamma^{(3)}_{ijk} :=
\rho_k \iN \gamma^{(4)}_{ij};
&  
\gamma^{(2)}_{ijkl} :=
\rho_l  \iN \gamma^{(3)}_{ijk}.
\nonumber
\end{array}
\end{equation}
 We have  also (using  $\delta^{ij}_{kl} := (\delta^i_k\delta^j_l- \delta^i_l\delta^j_k)$) the   useful relations:
\begin{equation}
\left\{
\begin{array}{rcl}
  {{\gamma}}^{i}\wedge {{\gamma}}_j^{(5)} &=& 
   \delta^i_j {{\gamma}}^{(6)},
\\
  {{\gamma}}^{i}\wedge {{\gamma}}_{kl}^{(4)} &=& 
   \delta^i_l {{\gamma}}^{(5)}_{k} -   \delta^i_k {{\gamma}}^{(5)}_{l},
\\
  \gamma^{ij} \wedge \gamma_{kl}^{(4)}  & =& \delta^{ij}_{kl}\gamma^{(6)}  ,
  \\
    \gamma^{ij} \wedge \gamma_{klm}^{(3)}  & =& \delta^{ij}_{lm}\gamma^{(5)}  +  \delta^{ij}_{mk}\gamma^{(5)}  +  \delta^{ij}_{kl}\gamma^{(5)}  .
  \nonumber \end{array}
  \right.
\end{equation} 
 \subsection{Einstein and Spin  forms}\label{Einstein-Spin}

\begin{lemm}\label{lem:algebraic-lemma-25}
{\em $\forall g\in \G$,  
$ {{\varepsilon_{abc}}}^{d}  (g{}^{-1}){}^{b}_{b'}   (g{}^{-1}){}^{c}_{c'}  g{}^{d'}_{d}   =   g{}^{a'}_{a} {{\varepsilon_{a'b'c'}}}^{d'}
$.  
  }
\end{lemm}
\noindent \Proof.   By definition of the Lorentz group we have,  $\forall g\in \G$,   $g{}_{a}{}^{b}  {\hbox{\sffamily h}}_{a'b'}   g{}^{b'}{}_{b}  = {\hbox{\sffamily h}}_{ab}  $ so that $\forall g \in \G$, $ (g{}^{-1}){}_{a}{}^{b} = {\hbox{\sffamily h}}_{aa'}  g{}^{a'}{}_{b'}  {\hbox{\sffamily h}}^{b'b} $. Note that   $\forall g\in \G$,  $\hbox{det} (g) = 1  $, then:
\begin{equation}
\begin{array}{rclcccl}
{{\varepsilon_{abcd}}}    &= &   {{\varepsilon_{a'b'c'd'}}}      g{}^{a'}_{a}  g{}^{b'}_{b}  g{}^{c'}_{c}g{}^{d'}_{d}    &\ \iff \ &  {{\varepsilon_{abc}}}^{d''}   &= & {{\varepsilon_{a'b'c'}}}^{d'}    g{}^{a'}_{a}  g{}^{b'}_{b}  g{}^{c'}_{c}         g{}^{d'''}_{d''}  {\hbox{\sffamily h}}_{d'''d'}    ,
 \\
&&   &\ \iff \ &       {{\varepsilon_{abc}}}^{d}
 &= & {{\varepsilon_{a'b'c'}}}^{d'}    g{}^{a'}_{a}  g{}^{b'}_{b}  g{}^{c'}_{c}      {\hbox{\sffamily h}}^{dd''}  g{}^{d'''}_{d''}  {\hbox{\sffamily h}}_{d'''d'} ,  
 \\
&&& \iff & {{\varepsilon_{abc}}}^{d}     & =&     {{\varepsilon_{a'b'c'}}}^{d'}    g{}^{a'}_{a}  g{}^{b'}_{b}  g{}^{c'}_{c}   ({g^{-1}}){}^{d}_{d'}.
        \end{array}
       \nonumber
\end{equation}
  \hfill $\square$

\subsubsection{Einstein $3$-forms}

 \begin{defi}\label{Einstein-3-form} 
The      Einstein  ${\mathfrak{t}}^*$-valued  
  $3$-form is denoted by ${{\Upsilon}} = {{\Upsilon}}_{a} \otimes {{t}}^{a} $,     where     $\forall 0\leq a \leq 3$,  ${{\Upsilon}}_{a} := 
 \frac{1}{2} 
 {{\epsilon}_{abc}}^{d}      (d \etao{}^{c}{}_d +    \etao{}^{c}{}_{c'} \wedge  \etao{}^{c'}{}_{d}  ) 
\wedge  \etaoo{}^{b}  $.
\end{defi} 
 
   \begin{lemm}\label{lem:algebraic-einstein}{\em
  $\forall \x\in {\cal X},\forall g\in \G$, if 
  $ \etao{}^{a}_{b}{}_\mu(x,g) = (g^{-1})^{a}_{b'} {A}_{\mu a'}^{b'}(x)g^{a'}_b $  and $\etaoo{}^{a}_{\mu} (x,g) = (g^{-1})^{a}_{a'} e_{\mu}^{a'} (x)$,
   then 
 $
  {{\Upsilon}}_a 
  = {{G}}_{a'} g^{a'}_{a} 
 $, where 
${{G}}_{a} =  \frac{1}{2} 
{{{\epsilon}_{abc}}}^{d}      \left(d {A}{}^{c}{}_{d} + {A}{}^{c}{}_{d'} \wedge {A{}^{d'}{}_{d}} \right) \wedge   {e}{}^{b} .
$
  }
 \end{lemm}

\noindent \Proof.   By using    Lemma \ref{lem:algebraic-lemma-25},   we have the straightforward calculation: 
 \[\left.
 \begin{array}{ccl}
  {{\Upsilon}}_a     & = &  
 \frac{1}{2}   {\varepsilon_{abc}}^{d} (g^{-1})^{c}_{c'}  ({g}{}^{-1})^{b}_{b'}  g^{d'}_{d}    (d A + A \wedge A  ){}^{c'}_{d'}  \wedge   e{}^{b'} 
   \\
 & = &      
    \frac{1}{2}  g{}^{a'}_{a} {{\varepsilon_{a'b'c'}}}^{d'}           (d {A}{}^{c'}{}_{d'} + {A}{}^{c'}{}_{d''} \wedge {A{}^{d''}{}_{d'}}  ) 
           \wedge   e{}^{b'} . 
   \end{array}
   \right.
\]
 \noindent Then,  $  {{\Upsilon}}_a =  {{G}}{}_{a'}    g{}^{a'}_{a}  $. 
  \hfill $\square$

 \begin{lemm}\label{Einstein-3-form-calculation}
{\em 
 $G_a =  G^{a'}{_{a}} e_{a'}^{(3)}  = G^{a'}{_{a}} {\pmb{e}}{}^{\mu}_{a'} \beta_{\mu}^{(3)} $, where $G^{a}{}_{b} $ are the components of the   Einstein tensor and $ {\pmb{e}}{}^{\mu}_{a}   :=  \hbox{det}(e) e^{\mu}_{a} 
$.}
\end{lemm}
\noindent {\Proof.}   Note that 
$
  G_a =  \frac{1}{2}   \varepsilon_{abcd}   e^{b} \wedge   F^{cd}   
    =   \frac{1}{2}   \varepsilon_{abcd}   e^{d} \wedge   F^{bc}     =      \frac{1}{2}  F^{bc} \wedge e^{(1)}_{abc} =      \frac{1}{4}   F^{bc}{}_{b'c'}     (   e^{b'c'}  \wedge  e_{abc}^{(1)}    ) 
$. 
     By using    algebraic  relations    in   Section \ref{co-yoga},  
we have the straightforward calculation:
 \[
\left.
\begin{array}{rcl}
\displaystyle      G_a      & = &   
      \frac{1}{2}      \left(    F^{bc}{}_{bc}    
   e^{(3)}_{a} 
+
   F^{bc}{}_{ca}    
 e^{(3)}_{b} 
 +   F^{bc}{}_{ab}   e^{(3)}_{c}  \right)
=    \frac{1}{2}     \left(  F^{bc}{}_{ca}    +     F^{cb}{}_{ac}      \right)   e_b^{(3)} 
  +   \frac{1}{2}  \hbox{S}     e_a^{(3)}   , 
      \\
 &=&    - \textsf{h}^{bb'}  F^{c}{}_{b'ca}      e_b^{(3)} 
  +   \frac{1}{2}  \hbox{S}     e_a^{(3)}   
=
 - \left(           \hbox{Ric}^{b}{}_{a}     -   \frac{1}{2}   \hbox{S}  \right)   e_b^{(3)}  
   \end{array}
\right.
\]
 Then $ G_a  =    - G^{b}{_a}   e_{b}^{(3)} $.  
 \hfill $\square$

   \begin{lemm}\label{Einstein-002}
    {\em We have the identity $   \left( \frac{1}{3!} \right) 
   \epsilon^{\mu\nu\rho\sigma}  G_{a\nu\rho\sigma}   = 
    G^{b}{}_a   e_{b}^{\mu}  $.
 }
 \end{lemm}

\noindent \Proof. Consider   Lemma   \ref{Einstein-3-form-calculation} and since $e^{(3)}_{a} :=   {\pmb{e}}{}^{\mu}_{a} \beta_{\mu}^{(3)}  := \left( \left(    \frac{1}{3!}  \right) \epsilon_{abcd} \epsilon^{\mu\nu\rho\sigma}   e^{b}_{\nu} e^{c}_{\rho} e^{d}_{\sigma} \right)   \beta_{\mu}^{(3)} $, we obtain:
  \[  
  \begin{array}{rcl}
 G_a  & 
 = & 
  - G^{a'}{_{a}} e_{a'}^{(3)}   =  -  \left( \frac{1}{3!} \right)    \epsilon_{a'bcd}    \epsilon^{\mu\nu\rho\sigma}  G^{a'}{_{a}}   e^{b}_{\nu} e^{c}_{\rho} e^{d}_{\sigma}  \beta^{(3)}_{\mu} 
  \\ 
 & = &   -  \left( \frac{1}{3!} \right)  \epsilon_{a'bcd}    \epsilon^{a'' bcd}  G^{a'}{_{a}}   e_{a''}^{\mu} \beta^{(3)}_{\mu}     
=    \delta^{a''}_{a'} G^{a'}{_{a}}   e_{a''}^{\mu} \beta^{(3)}_{\mu} 
 =       G^{b}{_{a}}   e_{b}^{\mu}\beta^{(3)}_{\mu}  .
  \end{array}
 \]
Also, since 
$
   2 G_a    =  
{{{\epsilon}_{abc}}}^{d}    {F}{}^{c}{}_{d}  \wedge   {e}{}^{b}    =    \epsilon_{abcd}e^d   \wedge F^{bc}  =  \frac{1}{2}    \epsilon_{abcd}e^d_{\rho}     F^{bc}_{\mu\nu} \beta^{\rho\mu\nu} 
$, we then obtain 
$
      G_a  =  \frac{1}{3!} G_{a}{}_{\mu\nu\rho}  \beta^{\mu\nu\rho} =      \frac{1}{3!}     {\epsilon}^{\sigma\lambda\mu\nu}     {{G}}{}_{a}{}_{\lambda\mu\nu} \beta_{\sigma}^{(3)}   
$,  where ${G}_{a}{}_{\mu\nu\rho}  =  
\left( \frac{3!}{4}  \right) \epsilon_{abcd}e^d_{\rho}     F^{bc}_{\mu\nu}
$.        \hfill $\square$

\subsubsection{Spin  $3$-forms}

\begin{defi}\label{Spin-forms}
The     Spin    $\g^*$-valued $3$-form is  ${{\Sigma}} := 
{{\Sigma}}_{c}{}^{d} \otimes u_i^{c}{}_d \mathfrak{l}^i$,     where  $\forall 0\leq a,b\leq 3$,  
$ {{\Sigma}}{}_{c}{}^{d}  :=
 \frac{1}{2} \epsilon_{abc}{}^{d} (d \etaoo{}^{a} +   \etao{}^{a}{}_{a'} \wedge  {\etaoo{}^{a'}}   
 )\wedge   {\etaoo{}^{b}} 
$. 
\end{defi}
  \begin{lemm}\label{lem:algebraic-spin}{\em
 $\forall \x\in {\cal X},\forall g\in \G$, if 
  $ \etao{}^{a}_{b}{}_\mu(x,g) = (g^{-1})^{a}_{b'} {A}_{\mu a'}^{b'}(x)g^{a'}_b $  and $\etaoo{}^{a}_{\mu} (x,g) = (g^{-1})^{a}_{a'} e_{\mu}^{a'} (x)$,
   then 
  $ {{\Sigma}}{}_{c}{}^{d} 
=   {{H}}_{c'}{}^{d'} g^{c'}_{c} (g^{-1})_{d'}^{d}  , 
$ where
$ {{H}}_{c}{}^{d} =  \frac{1}{2}   {{{\epsilon}_{abc}}}^{d}    (d {e}{}^{a} + {A}{}^{a}{}_{a'} \wedge {e}{}^{a'} )   \wedge    {e}{}^{b} 
$.  
  }
 \end{lemm}

\noindent \Proof.   By using    Lemma \ref{lem:algebraic-lemma-25},   we have the straightforward calculation: 
 \[
 \begin{array}{rcl}
  {{\Sigma}}_{c}{}^{d} 
 &=&   \frac{1}{2}   {\varepsilon_{abc}}^{d} (g^{-1})^{a}_{a'}   ({g}{}^{-1})^{b}_{b'} \left(d e + A \wedge e \right){}^{a'}    \wedge    e{}^{b'}   
 \\
&=&  \frac{1}{2}   {\varepsilon_{a'b'c'}}^{d'} (g)^{c'}_{c}   ({g}{}^{-1})^{b}_{b'}  (d e^{a'} + A^{a'}{}_{a''} \wedge e^{a''}  )    \wedge    e{}^{b'} 
\end{array}
  \]
 \noindent Then,     $ {{\Sigma}}_{c}{}^{d}  = {{H}}{}_{c}{}^{d}     g{}^{c'}_{c} (g{}^{-1}){}^{d}_{d'} $. 
  \hfill $\square$

\begin{lemm}\label{Spin-3-form-calculation}
 {\em 
 $    {H}_{a}{}^{b}   =     \frac{1}{2}   {\textsf{h}}^{bb'}       (    T^{c}{}_{b'c}    
   e^{(3)}_{a} 
+
   T^{c}{}_{ca}    
 e^{(3)}_{b'} 
 -   T^{c}{}_{b'a}   e^{(3)}_{c}   )  $, where $    T^{c}{}_{ab}   $ are the components of the torsion  tensor.
 }
\end{lemm}
\noindent {\Proof.}    Note that 
$
  {H}_{a}{}^{b}   :=   \frac{1}{2}   \varepsilon_{ab'cd}   {\textsf{h}}^{bb'}  e^{d} \wedge   T^{c}   
 =   
 \frac{1}{2}   {\textsf{h}}^{bb'}    T^{c} \wedge e^{(1)}_{ab'c} =   \frac{1}{4}    {\textsf{h}}^{bb'}    T^{c}{}_{a'c'}     (   e^{a'c'}  \wedge  e_{ab'c}^{(1)}    )  
$. By using    algebraic  relations given   in   Section \ref{co-yoga}: 
\[
\left.
\begin{array}{rcl}
\displaystyle      {H}_{a}{}^{b}   & = &  
     \frac{1}{4}    {\textsf{h}}^{bb'}    T^{c}{}_{a'c'}     \left( \delta^{a'c'}_{b'c} e^{(3)}_{a} +
\delta^{a'c'}_{ca} e^{(3)}_{b} +
\delta^{a'c'}_{ab'} e^{(3)}_{c} \right)
 \\
 &=& 
   \frac{1}{2}   {\textsf{h}}^{bb'}      \left(    T^{c}{}_{b'c}    
   e^{(3)}_{a} 
+
   T^{c}{}_{ca}    
 e^{(3)}_{b'} 
 +  T^{c}{}_{ab'}   e^{(3)}_{c}  \right) .
    \end{array}
\right.
\]
\hfill $\square$
  
   \begin{lemm}\label{Spin-002}
    {\em The identity $ \left( \frac{1}{3!} \right) \epsilon^{\mu\nu\rho\sigma}  H_{a}{}^{b}{}_{\nu\rho\sigma}   =    \frac{1}{2}  {\textsf{h}}^{bb'}     \left(    T^{c}{}_{b'c}   
   \delta^{a'}_{a}      +     T^{c}{}_{ca}   \delta^{a'}_{b'}           
  +  T^{c}{}_{ab'}  \delta^{a'}_{c}  \right)    e_{a'}^{\mu}    $ holds.
 }
 \end{lemm}
 \noindent \Proof. By using Lemma \ref{Spin-3-form-calculation} and since $e^{(3)}_{a} :=   {\pmb{e}}{}^{\mu}_{a} \beta_{\mu}^{(3)}  := \left( \left(    \frac{1}{3!}  \right) \epsilon_{abcd} \epsilon^{\mu\nu\rho\sigma}   e^{b}_{\nu} e^{c}_{\rho} e^{d}_{\sigma} \right)   \beta_{\mu}^{(3)} $, we have: 
   \[
  \begin{array}{rcl}
   {H}_{a}{}^{b}  
&=& \frac{1}{2}  \left[ {\textsf{h}}^{bb'}      \left(    T^{c}{}_{b'c}    {\pmb{e}}{}^{\mu}_{a}+T^{c}{}_{ca}       {\pmb{e}}{}^{\mu}_{b'}+  T^{c}{}_{ab'}      {\pmb{e}}{}^{\mu}_{c}   \right)   \right]  \beta_{\mu}^{(3)} 
   \\
     & =&  
    \left( \frac{1}{2 \cdot 3!} \right)  \left[ \epsilon^{\mu\nu\rho\sigma}     {\textsf{h}}^{bb'}       \left(   \epsilon_{adef}      T^{c}{}_{b'c}    
     +
  \epsilon_{b'def}    T^{c}{}_{ca}     
  +    \epsilon_{cdef}  T^{c}{}_{ab'}     
 \right)    e^{d}_{\nu}     e^{e}_{\rho}    e^{f}_{\sigma}   \right]  \beta_{\mu}^{(3)} 
  \\
   & =& -  \frac{1}{2  }     \left(  \delta^{a'}_{a}    T^{c}{}_{b'c}     e_{a'}^{\mu}            + T^{c}{}_{ca}    \delta_{b'}^{b''}   e_{b''}^{\mu}         +   \delta_{c'}^{c}  T^{c}{}_{ab'}     e_{c'}^{\mu}        \right)  \beta_{\mu}^{(3)} 
\\
      &=  & - 
   \frac{1}{2}  {\textsf{h}}^{bb'}     \left(    T^{c}{}_{b'c}   
   \delta^{a'}_{a}      +     T^{c}{}_{ca}   \delta^{a'}_{b'}           
  +  T^{c}{}_{ab'}  \delta^{a'}_{c}  \right)    e_{a'}^{\mu}         \beta_{\mu}^{(3)} .
  \end{array}
 \]
 where $ e_{a'}^{\mu} $ is such that $e_{a'}^{\mu} (x)  e^{a}_{\mu} (x)= \delta^{a}_{a'}$.
 Finally,   since  
 $
 2 H_{a}{}^{b} 
=    \epsilon_{cda}{}^{b}  e^d   \wedge T^{c} 
 =       \epsilon_{cdab'}{} {\textsf{h}}^{bb'}  e^d   \wedge T^{c}   
 = \frac{1}{2}  \epsilon_{ab'cd}e^d_{\rho} {\textsf{h}}^{bb'}  T^{c}_{\mu\nu} \beta^{\rho\mu\nu} ,
 $, then    the    Spin $3$-forms $ {H}_{c}{}^{d}$   are given by 
$
  {H}_{a}{}^{b}  =  \frac{1}{3!}    {H}_{a}{}^{b}{}_{\mu\nu\rho}  \beta^{\mu\nu\rho} =      \frac{1}{3!}     {\epsilon}^{\sigma\lambda\mu\nu}     {H}_{a}{}^{b}{}_{\lambda\mu\nu} \beta_{\sigma}^{(3)}   
$, where $  {H}_{a}{}^{b}{}_{\mu\nu\rho}  = \left( \frac{3!}{4}  \right)   \epsilon_{ab'cd}e^d_{\rho}     T^{c}_{\mu\nu}$. 
       \hfill $\square$

 \subsection{Coadjoint exterior yoga}\label{coadjointyoga}
  
  Let us denote by ${\cal E}_{abc}{}^{d} := {\etao{}^{a'}_{a}}   {\varepsilon_{a'bc}}^{d}  +    {\etao{}^{b'}_{b}}   {\varepsilon_{ab'c}}^{d}   +   {\etao{}^{c'}_{c}}   {\varepsilon_{abc'}}^{d}   -  {\etao{}^{d}_{d'}}   {\varepsilon_{abc}}^{d'}    $
 \begin{lemm}\label{algebraic-lemma-01}
{\em  The identity   ${\cal E}_{abc}{}^{d} = 0$ holds. Equivalently,    
$          {\etao{}^{d}_{d'}}   {\varepsilon_{abc}}^{d'} -  {\etao{}^{c'}_{c}}   {\varepsilon_{abc'}}^{d}   =   {\etao{}^{a'}_{a}}   {\varepsilon_{a'bc}}^{d}  +    {\etao{}^{b'}_{b}}   {\varepsilon_{ab'c}}^{d} $. 
}
\end{lemm}
\noindent \Proof. 
 $
  -    {\etao{}^{d}_{d'}}   {\varepsilon_{abc}}^{d'}   
    =  {\varepsilon_{abcd''}}  (  -    {\etao{}^{d}_{d'}}  {\hbox{{\sffamily h}}}^{d'd''} ) =   {\varepsilon_{abcd''}}  (   {\hbox{{\sffamily h}}}^{dd'}  {\etao{}^{d''}_{d'}}  )          =    {\varepsilon_{abcd''}}      {\etao{}^{d''}_{d'}}  {\hbox{{\sffamily h}}}^{d'd}        
$.   Therefore:
 ${\cal E}_{abc}{}^{d} 
 =   (  {\etao{}^{a'}_{a}}   {\varepsilon_{a'bcd''}}   +    {\etao{}^{b'}_{b}}   {\varepsilon_{ab'cd''}}    +   {\etao{}^{c'}_{c}}   {\varepsilon_{abc'd''}}  +   {\etao{}^{d''}_{d'}}   {\varepsilon_{abcd''}}  )  {\hbox{{\sffamily h}}}^{d'd}     ,
 $ which is identically vanishing.
           \hfill $\square$

           \begin{lemm} {\em
           On the  submanifold of constraints ${\cal N}$, see \eqref{constraint}, we have 
           \[
            {\hbox{ad}}_{\eta_\nu}^*({{\psi}}{}^{\mu\nu})  =  \frac{1}{2} [\nbetaoo{}^{\mu\nu},\etao_\nu]{}^{d}_{c} u^{c}{}_d
            \quad \quad {\hbox{ad}}_{{\textswab{l}_{j}}}^*({{\psi}}{}^{\mu j}) =   \frac{1}{2} [\bfpo{}^{\mu j}, \mathfrak{l}_j]{}^{d}_{c} u^{c}{}_d+ \left( \bfpoo{}_{c'}^{\mu j}    {{\textswab{l}_{j}}{}^{c'}_{c}} \right) t^{c}
            \]
           }
           \end{lemm}

\noindent \Proof.  First, using  Equation \eqref{coadjoint-dual} in Section \ref{lie-rep}, 
  \[\left\{
  \begin{array}{rcl}
  {\hbox{ad}}_{\eta_\nu}^*({{\psi}}{}^{\mu\nu})  &=&
  \frac{1}{2} \left( {\bfpo{}^{d}_{c'}}{}^{\mu\nu}     {\etao{}^{c'}_{c}}_{\nu}      -  {\etao{}^{d}_{d'}}_{\nu}   {\bfpo{}^{d'}_{c}}{}^{\mu\nu}  - 2 \bfpoo{}_c^{\mu\nu}     \etaoo{}^{d}_{\nu} \right) u^{c}{}_{d}
   +  \left( \bfpoo{}_c^{\mu\nu} \etaoo{}^{d}_\mu  \right)   t^{c}
   \\
  {\hbox{ad}}_{{\textswab{l}_{j}}}^*({{\psi}}^{\mu j})     &=&
  \frac{1}{2}     \left(  {\bfpo{}^{d}_{c'}}{}^{\mu j}   {{\textswab{l}_{j}}{}^{c'}_{c}}     -  {{\textswab{l}_{j}}{}^{d}_{d'}}    {\bfpo{}^{d'}_{c}}{}^{\mu j}  
  \right)    u^{c}{}_{d}  + \left(   {\bfpoo{}_{c'}}{}^{\mu j}    {{\textswab{l}_{j}}{}^{c'}_{c}}   \right)    {{t}}^{c}  
  \end{array}
  \right.
 \]
 We conclude by using relations \eqref{correspondanceLegendre}  {i.e.}  $ \po{}^d_{c}{}^{\mu\nu}   =  {{\nbetaoo}}{}^{d\mu\nu}_{c}$ and $      \poo{}_{a}^{\mu\nu} =  0$, where we denote $   [\nbetaoo{}^{\mu\nu},\etao_\nu]{}^{d}_{c} :=   {{\nbetaoo}}{}^{d\mu\nu}_{c'}    {\etao{}^{c'}_{c}}_{\nu}      -  {\etao{}^{d}_{d'}}_{\nu}   {{\nbetaoo}}{}^{d'\mu\nu}_{c}       $ and $   [\bfpo{}^{\mu j}, \mathfrak{l}_j]  :=  {\bfpo{}^{d}_{c'}}{}^{\mu j}   {{\textswab{l}_{j}}{}^{c'}_{c}}     -  {{\textswab{l}_{j}}{}^{d}_{d'}}    {\bfpo{}^{d'}_{c}}{}^{\mu j}    $.
    \hfill $\square$

\begin{lemm}\label{lem:constcalcul1} {\em
On    ${\cal N} \subset {\cal N}$ (see \eqref{constraint}), we have 
$( \hbox{ad}_{\eta_\nu}^*({{\psi}}^{\mu\nu})  ){}^{d}_{c}  \beta^{(3)}_{\mu}  =     {\varepsilon_{abc}}^{d}  {\etao{}^{a}_{a'}} \wedge  {\etaoo{}^{a'}}     \wedge   {\etaoo{}^{b}}      $.
}
\end{lemm}
\noindent \Proof. Consider the $2$-form $\bfpo = \frac{1}{2} \bfpo{}^{\mu\nu} \beta^{(2)}_{\mu\nu}$ and the $1$-form $\etao = \etao{}_{\rho}^{} dx^{\rho}$, Then: 
\[\begin{array}{ccl}
[{\bfpo{}}  \wedge  {\etao{}} ] &:=&   \frac{1}{2} \left( {\bfpo{}}{}^{\mu\nu}   {\etao{}}{}_\rho  \beta_{\mu\nu}^{(2)} \wedge   dx^{\rho}       -      {\etao{}}_{\rho}   {\bfpo{}}{}^{\mu\nu}    dx^{\rho}  \wedge  \beta_{\mu\nu}^{(2)} \right) 
 \\
 &=& 
 \frac{1}{2} \left(  {\bfpo{}}{}^{\mu\nu}   {\etao{}}{}_\rho  \left(\delta^{\rho}_{\nu} \beta_{\mu}^{(3)} - \delta^{\rho}_{\mu} \beta_{\nu}^{(3)} \right)        -      {\etao{}}_{\rho}    {\bfpo{}}{}^{\mu\nu}    \left(\delta^{\rho}_{\nu} \beta_{\mu}^{(3)} - \delta^{\rho}_{\mu} \beta_{\nu}^{(3)} \right)   \right)
\\
 & = &   {\bfpo{}}{}^{\mu\nu}    {\etao{}}_{\nu} \beta_{\mu}^{(3)}   -    {\etao{}}_{\nu}    {\bfpo{}}{}^{\mu\nu} \beta_\mu^{(3)}   
  \\
   & = &
        [ {\bfpo{}}{}^{\mu\nu}  ,    {\etao{}}_{\nu}  ] \beta_{\mu}^{(3)} .      \\
\end{array}
\]

\noindent Finally, on the   submanifold  of constraints ${\cal N} $, we have $\po{}^d_{c}{}^{\mu\nu}    =    {{\nbetaoo}}{}^{d\mu\nu}_{c} $ (see   \eqref{correspondanceLegendre}), therefore, we  obtain             $\frac{1}{2} {\bfpo{}^{d}_{c}}{}^{\mu\nu}  \beta_{\mu\nu}^{(2)} = \frac{1}{2} {\varepsilon_{abc}}^{d}     {\etaoo{}^{ab}}      $, where        ${\etaoo{}^{ab}}   :=  {\etaoo{}^{a}}     \wedge   {\etaoo{}^{b}}    $.
\[ 
\begin{array}{ccl}
   ( \hbox{ad}_{\eta_\nu}^*({{\psi}}^{\mu\nu})  ){}^{d}_{c}   \beta_{\mu}^{(3)}       &=& 
           \frac{1}{2}  \left(  {\varepsilon_{abc'}}^{d}     {\etaoo{}^{ab}}      \wedge  {\etao{}^{c'}_{c}}   -   {\varepsilon_{abc}}^{d'}     {\etaoo{}^{ab}}     \wedge    {\etao{}^{d}_{d'}}  \right)      
  \\
   & = &
      \frac{1}{2}  \left(   {\etao{}^{c'}_{c}}     {\varepsilon_{abc'}}^{d}   -   {\etao{}^{d}_{d'}}      {\varepsilon_{abc}}^{d'}   \right) \wedge  {\etaoo{}^{ab}}  
           \\
   & = &  
     \frac{1}{2} \left(   {\varepsilon_{a'bc}}^{d}  {\etao{}^{a'}_{a}} \wedge  {\etaoo{}^{ab}}     +        {\varepsilon_{b'ac}}^{d}  {\etao{}^{b'}_{b}} \wedge    {\etaoo{}^{ba}}   \right)  
  \\ & =&
      {\varepsilon_{abc}}^{d}  {\etao{}^{a}_{a'}} \wedge  {\etaoo{}^{a'b}}      ,  
        \end{array}
        \]
        where   in the second line we have used   Lemma \ref{algebraic-lemma-01}.
  \hfill $\square$
  
 \begin{lemm}\label{lemma-alg} {\em 
$\forall g\in \G$,   $ \Xio{}_{c}^{d}{}^{\mu}  $ and $ \Xioo{}_{a}{}^{\mu}$ are given by:
  \begin{equation}
    \Xio{}_{c}^{d}{}^{\mu}  
    = ({g}{}^{-1})^{d}_{d'} \left(  {\overset{\mathfrak{0}}{{{p}}}{}^{d'}_{c'}}{}^{\mu j}_{;j}  \right)    g^{c'}_{c}    ,  
 \quad \quad  
  \Xioo{}_{a}{}^{\mu}     
 =
 \left(    \overset{\mathfrak{1}}{{{p}}}{}^{\mu j}_{a';j}  \right)      g^{a'}_{a}       ,  
 \end{equation}
where   $ {\overset{\mathfrak{0}}{{{p}}}{}^{d}_{c}}{}^{\mu j} $ and $\overset{\mathfrak{1}}{{{p}}}{}^{\mu j}_{a}$ are given in Proposition    \ref{HVDW-002}.}
 \end{lemm}
\noindent  \Proof. Note that $\forall g\in \G$,  $(g^{-1}dg)^{i}$ is the component of  the Maurer--Cartan $1$-form in the basis ${\mathfrak{l}}_{i}$.  Note also that     
  $ d  g^{-1} = - g^{-1}  d  g g^{-1}$. Straightforwardly: 
\begin{equation}
\left\{
\begin{array}{rcl}
    \Xio{}_{c}^{d}{}^{\mu}  
      & = & \displaystyle   {\bfpo{}^{d}_{c}}{}_{;j}{}^{\mu j}  + [ {\textswab{l}}_{j} ,      {\bfpo}{}^{\mu j} ]{}^{d}_{c}    
\\&=&
  {\bfpo{}^{d}_{c}}{}^{\mu j}{}_{;j}       + 
   \left(  (g{}^{-1})^{d}_{d'} 
g^{d'}_{d''}{}_{;j}  \right)  {\bfpo{}^{d''}_{c}}{}^{\mu j}    
-      {\bfpo{}^{d}_{c''}}{}^{\mu j}  \left(    (g{}^{-1})^{c''}_{b} g^{b}_{c}{}_{;j}     \right) 
 \\
& =&    (g{}^{-1})^{d}_{d'} 
\left(g^{d'}_{d''}{}_{;j}   {\bfpo{}^{d''}_{c''}}{}^{\mu j}     (g{}^{-1})^{c''}_{c'}
+ g^{d'}_{d''}   {\bfpo{}^{d''}_{c''}}{}^{\mu j}{}_{;j}     (g{}^{-1})^{c''}_{c'} \right.
 \\
&  & 
\hfill \left. - g^{d'}_{d''}   {\bfpo{}^{d''}_{c''}}{}^{\mu j}     (g{}^{-1})^{c''}_{b} g^{b}_{b'}{}_{;j} (g{}^{-1})^{b'}_{c'} 
   \right)
 g^{c'}_{c}      
\\
& =&    (g{}^{-1})^{d}_{d'} (g^{d'}_{d''}   {\bfpo{}^{d''}_{c''}}{}^{\mu j}     (g{}^{-1})^{c''}_{c'}   )_{;j} g^{c'}_{c}  , 
\\
  \overset{\mathfrak{1}}{{{{\Xi}}}}{}_{a}{}^{\sigma}     
  & =  &     {\bfpoo{}_{a}}{}_{;j}^{\mu j}  -   {\bfpoo{}_{b}}{}^{\mu j}    {{\textswab{l}_{j}}{}^{b}_{a}}  
       \\
       & = &      
      {\bfpoo{}_{a}}{}_{;j}^{\mu j}  -   {\bfpoo{}_{a''}}{}^{\mu j}   (g^{-1})^{a''}_{b} g^{b}_{a}{}_{;j}        \\
       & = & \displaystyle        \left(
     {\bfpoo{}_{a''}}{}_{;j}^{\mu j}     (g^{-1})^{a''}_{a'}   -   {\bfpoo{}_{a''}}{}^{\mu j}   (g^{-1}){}^{a''}_{b} g{}^{b}_{b'}{}_{;j}  (g^{-1})^{b'}_{a'} \right)  g^{a'}_{a}      
   \\
      & = &    
       (  
     {\bfpoo{}_{a''}}{}^{\mu j}    (g^{-1})^{a''}_{a'}    )_{;j}    g^{a'}_{a}     . 
   \end{array}
   \right.
   \nonumber
\end{equation}
 Then, $    \Xio{}_{c}^{d}{}^{\mu}  =   ({g}{}^{-1})^{d}_{d'} \left(  {\overset{\mathfrak{0}}{{{p}}}{}^{d'}_{c'}}{}^{\mu j}_{;j}  \right)    g^{c'}_{c}   $ and  $   \overset{\mathfrak{1}}{{{{\Xi}}}}{}_{a}{}^{\sigma}      =  \left(    \overset{\mathfrak{1}}{{{p}}}{}^{\mu j}_{a';j}  \right)      g^{a'}_{a}  $, respectively.  \hfill $\square$



\end{document}